\documentclass[twocolumn,astrosym]{aastex61}
\usepackage{CJK}

\shorttitle{LaCoSSPAr in the SGC}
\shortauthors{Yang et al.}

\begin{document}
\begin{CJK*}{UTF8}{gbsn}

\title{The LAMOST Complete Spectroscopic Survey of Pointing Area (LaCoSSPAr)\\
in the Southern Galactic Cap\\
I. The Spectroscopic Redshift Catalog}

\author{Ming Yang (杨明)}
\altaffiliation{LAMOST Fellow, National Astronomical Observatories, Chinese Academy of Sciences}
\affiliation{Key Laboratory of Optical Astronomy, National Astronomical Observatories, Chinese Academy of Sciences, Datun Road 20A, Beijing 100012, China}
\affiliation{IAASARS, National Observatory of Athens, Vas. Pavlou \& I. Metaxa, Penteli 15236, Greece}
\author{Hong Wu (吴宏)}
\affiliation{Key Laboratory of Optical Astronomy, National Astronomical Observatories, Chinese Academy of Sciences, Datun Road 20A, Beijing 100012, China}
\author{Fan Yang (杨帆)}
\affiliation{Key Laboratory of Optical Astronomy, National Astronomical Observatories, Chinese Academy of Sciences, Datun Road 20A, Beijing 100012, China}
\author{Man I Lam (林敏仪)}
\affiliation{Key Laboratory for Research in Galaxies and Cosmology, Shanghai Astronomical Observatory, Chinese Academy of Sciences, 80 Nandan Road, Shanghai 200030, China}
\author{Tian-Wen Cao (曹天文)}
\affiliation{Key Laboratory of Optical Astronomy, National Astronomical Observatories, Chinese Academy of Sciences, Datun Road 20A, Beijing 100012, China}
\author{Chao-Jian Wu (武朝剑)}
\affiliation{Key Laboratory of Optical Astronomy, National Astronomical Observatories, Chinese Academy of Sciences, Datun Road 20A, Beijing 100012, China}
\author{Pin-Song Zhao (赵品松)}
\affiliation{Key Laboratory of Optical Astronomy, National Astronomical Observatories, Chinese Academy of Sciences, Datun Road 20A, Beijing 100012, China}
\author{Tian-Meng Zhang (张天萌)}
\affiliation{Key Laboratory of Optical Astronomy, National Astronomical Observatories, Chinese Academy of Sciences, Datun Road 20A, Beijing 100012, China}
\author{Zhi-Min Zhou (周志民)}
\affiliation{Key Laboratory of Optical Astronomy, National Astronomical Observatories, Chinese Academy of Sciences, Datun Road 20A, Beijing 100012, China}
\author{Xue-Bing Wu (吴学兵)}
\affiliation{Department of Astronomy, School of Physics, Peking University, Beijing 100871, China}
\author{Yan-Xia Zhang (张彦霞)}
\affiliation{Key Laboratory of Optical Astronomy, National Astronomical Observatories, Chinese Academy of Sciences, Datun Road 20A, Beijing 100012, China}
\author{Zheng-Yi Shao (邵正义)}
\affiliation{Key Laboratory for Research in Galaxies and Cosmology, Shanghai Astronomical Observatory, Chinese Academy of Sciences, 80 Nandan Road, Shanghai 200030, China}
\author{Yi-Peng Jing (景益鹏)}
\affiliation{School of Physics and Astronomy, Shanghai Jiao Tong University, 800 Dongchuan Road, Shanghai 200240, China}
\author{Shi-Yin Shen (沈世银)}
\affiliation{Key Laboratory for Research in Galaxies and Cosmology, Shanghai Astronomical Observatory, Chinese Academy of Sciences, 80 Nandan Road, Shanghai 200030, China}
\author{Yi-Nan Zhu (朱轶楠)}
\affiliation{Key Laboratory of Optical Astronomy, National Astronomical Observatories, Chinese Academy of Sciences, Datun Road 20A, Beijing 100012, China}
\author{Wei Du (杜薇)}
\affiliation{Key Laboratory of Optical Astronomy, National Astronomical Observatories, Chinese Academy of Sciences, Datun Road 20A, Beijing 100012, China}
\author{Feng-Jie Lei (雷凤杰)}
\affiliation{Key Laboratory of Optical Astronomy, National Astronomical Observatories, Chinese Academy of Sciences, Datun Road 20A, Beijing 100012, China}
\author{Min He (何敏)}
\affiliation{Key Laboratory of Optical Astronomy, National Astronomical Observatories, Chinese Academy of Sciences, Datun Road 20A, Beijing 100012, China}
\author{Jun-Jie Jin (金骏杰)}
\affiliation{Key Laboratory of Optical Astronomy, National Astronomical Observatories, Chinese Academy of Sciences, Datun Road 20A, Beijing 100012, China}
\author{Jian-Rong Shi (施建荣)}
\affiliation{Key Laboratory of Optical Astronomy, National Astronomical Observatories, Chinese Academy of Sciences, Datun Road 20A, Beijing 100012, China}
\author{Wei Zhang (张伟)}
\affiliation{Key Laboratory of Optical Astronomy, National Astronomical Observatories, Chinese Academy of Sciences, Datun Road 20A, Beijing 100012, China}
\author{Jian-Ling Wang (王建岭)}
\affiliation{Key Laboratory of Optical Astronomy, National Astronomical Observatories, Chinese Academy of Sciences, Datun Road 20A, Beijing 100012, China}
\author{Yu-Zhong Wu (吴玉中)}
\affiliation{Key Laboratory of Optical Astronomy, National Astronomical Observatories, Chinese Academy of Sciences, Datun Road 20A, Beijing 100012, China}
\author{Hao-Tong Zhang (张昊彤)}
\affiliation{Key Laboratory of Optical Astronomy, National Astronomical Observatories, Chinese Academy of Sciences, Datun Road 20A, Beijing 100012, China}
\author{A-Li Luo (罗阿里)}
\affiliation{Key Laboratory of Optical Astronomy, National Astronomical Observatories, Chinese Academy of Sciences, Datun Road 20A, Beijing 100012, China}
\author{Hai-Long Yuan (袁海龙)}
\affiliation{Key Laboratory of Optical Astronomy, National Astronomical Observatories, Chinese Academy of Sciences, Datun Road 20A, Beijing 100012, China}
\author{Zhong-Rui Bai (白仲瑞)}
\affiliation{Key Laboratory of Optical Astronomy, National Astronomical Observatories, Chinese Academy of Sciences, Datun Road 20A, Beijing 100012, China}
\author{Xu Kong (孔旭)} 
\affiliation{Center for Astrophysics, University of Science and Technology of China, Hefei 230026, China}
\author{Qiu-Sheng Gu (顾秋生)} 
\affiliation{Department of Astronomy, Nanjing University, Nanjing 210093, China}
\author{Yong Zhang (张勇)}
\affiliation{Nanjing Institute of Astronomical Optics \& Technology, National Astronomical Observatories, Chinese Academy of Sciences, Nanjing 210042, China}
\author{Yong-Hui Hou (侯永辉)}
\affiliation{Nanjing Institute of Astronomical Optics \& Technology, National Astronomical Observatories, Chinese Academy of Sciences, Nanjing 210042, China}
\author{Yong-Heng Zhao (赵永恒)}
\affiliation{Key Laboratory of Optical Astronomy, National Astronomical Observatories, Chinese Academy of Sciences, Datun Road 20A, Beijing 100012, China}

\email{myang@nao.cas.cn, hwu@nao.cas.cn}

\begin{abstract}
We present a spectroscopic redshift catalog from the LAMOST Complete Spectroscopic Survey of Pointing Area (LaCoSSPAr) in the Southern Galactic Cap (SGC), which is designed to observe all sources (Galactic and extra-galactic) by using repeating observations with a limiting magnitude of $r=18.1~mag$ in two $20~deg^2$ fields. The project is mainly focusing on the completeness of LAMOST ExtraGAlactic Surveys (LEGAS) in the SGC, the deficiencies of source selection methods and the basic performance parameters of LAMOST telescope. In both fields, more than 95\% of galaxies have been observed. A post-processing has been applied to LAMOST 1D spectrum to remove the majority of remaining sky background residuals. More than 10,000 spectra have been visually inspected to measure the redshift by using combinations of different emission/absorption features with uncertainty of $\sigma_{z}/(1+z)<0.001$. In total, there are 1528 redshifts (623 absorption and 905 emission line galaxies) in Field A and 1570 redshifts (569 absorption and 1001 emission line galaxies) in Field B have been measured. The results show that it is possible to derive redshift from low SNR galaxies with our post-processing and visual inspection. Our analysis also indicates that up to 1/4 of the input targets for a typical extra-galactic spectroscopic survey might be unreliable. The multi-wavelength data analysis shows that the majority of mid-infrared-detected absorption (91.3\%) and emission line galaxies (93.3\%) can be well separated by an empirical criterion of $W2-W3=2.4$. Meanwhile, a fainter sequence paralleled to the main population of galaxies has been witnessed both in $M_r$/$W2-W3$ and $M_*$/$W2-W3$ diagrams, which could be the population of luminous dwarf galaxies but contaminated by the edge-on/highly inclined galaxies ($\sim30\%$). 
\end{abstract}


\section{Introduction}

Redshift is the fundamental parameter for the extra-galactic objects. With spectroscopic redshift, the accurate distance information can be obtained to reveal the intrinsic properties of those objects, e.g., the clustering of galaxies which can shed light on the important information about large-scale structure and cosmological models \citep{Bahcall1988, Postman1992, Voit2005, Diaferio2008, Wen2010}, the pair counts and fraction which are the indicators of galaxy merger rate \citep{Burkey1994, Le2000, Kitzbichler2008, Lotz2011}, the velocity dispersion of galaxies in groups and clusters which can be used to deduce the dark matter halo \citep{Carlberg1996, Jing1998, Springel2001, Berlind2006, Thomas2011}, the external environment of field/cluster galaxies which is related to the formation and evolution of galaxies \citep{Balogh1999, Bell2004, Kauffmann2004, Elbaz2007, Peng2010}, the galaxy population which is classified by the fundamental properties such as color, luminosity, metallicity, stellar mass ($M_*$), star formation rate (SFR), surface brightness and so on \citep{Faber1976, Kennicutt1998, Cole2001, Baldry2004, Bower2006, Kewley2006}. Meanwhile, plenty of information about stellar population, emission line strength/ratio, SFR, metallicity and active galactic nucleus (AGN) activities also can be diagnosed by the spectrum itself \citep{Baldwin1981, Kennicutt1983, Sanders1988, Zaritsky1994, Calzetti2000, Bruzual2003}. Finally, spectroscopic redshift can be used in combination with multi-wavelength data to find rare and interesting objects, or served as the primary standard to calibrate the photometric redshifts.

Based on the advances in technology over the past couple of decades, many spectroscopic surveys, which obtain thousands to millions of redshifts, have been conducted to map the universe from small- to large-scales. The prominent ones among them include, early CfA and CfA2 Redshift Survey \citep{Huchra1983, Falco1999}, the Two-Degree Field (2dF) Galaxy Redshift Survey \citep{Lewis2002}, the Sloan Digital Sky Survey (SDSS) \citep{York2000, Dawson2013}, the Galaxy And Mass Assembly (GAMA) redshift survey \citep{Baldry2010} and so on. Although plenty of important achievements have been made by these surveys, obtaining more and accurate spectroscopic redshifts in different redshift ranges is still an important task for the better understanding of many topics (e.g. the physical properties of dark energy and dark matter, the galaxy formation and evolution at different redshifts, the evolution of luminosity function and clustering of galaxies).

The Large Sky Area Multi-Object Fiber Spectroscopic Telescope (LAMOST, also known as the Guoshoujing telescope, GSJT, \citealt{Wang1996, Su2004, Cui2012, Zhao2012}), as one of the National Major Scientific Projects, is a Wang-Su Schmidt telescope located in Xinglong Station of National Astronomical Observatory, Chinese Academy of Sciences (NAOC). With a 4-meter effective aperture, $\sim20~deg^2$ field of view (FOV, a diameter of $5\arcdeg$) and 4000 optical fibers (an aperture of 3.3" for each fiber), LAMOST is able to simultaneously observe more than 3000 scientific targets in a single exposure with designed limiting magnitude of $r=18.0$ mag and cover the wavelength range of $3700\sim9000 \textup{\AA}$ ($3700\sim5900\textup{\AA}$ for the blue and $5700\sim9000\textup{\AA}$ for the red spectrograph arms, respectively) with resolution of $R\approx1800$. Thus, LAMOST has a great potential to efficiently survey a large sky area for stars and galaxies. 

During the period from Oct 2011 to June 2012, LAMOST conducted a pilot survey to demonstrate the instrumental performance and feasibility of the science goals. The large-scale scientific survey covering the sky area of $-10<\delta<60$ started from September 2012. Currently (at the time of writing), LAMOST is conducting the fifth year of scientific survey begins on September 2016. Also, the LAMOST Data Release 4 (DR4) has been internal released with 7,681,185 spectra and 3,454 plates, including 6,898,298 stars, 118,743 galaxies and 41,352 quasars, from the pilot and first four-years scientific survey. Among them, the number of targets with signal-to-noise ratio (SNR) greater than 10 in g- or i-band is over six million. For the five-year plan of LAMOST, it may release more than nine million spectra of different kinds of targets. 

As an important part of LAMOST scientific survey, LAMOST ExtraGAlactic Surveys (LEGAS) aims to take hundreds of thousands of spectra for extra-galactic objects over $8000~deg^2$ of the Northern Galactic Cap (NGC) and $3500~deg^2$ of the Southern Galactic Cap (SGC) in five years. Involving in the survey, we conduct a LAMOST Key Project, named as the LAMOST Complete Spectroscopic Survey of Pointing Area (LaCoSSPAr) in the SGC, which is designed to observe all sources (Galactic and extra-galactic) by using repeating observational strategy with a limiting magnitude of $r=18.1~mag$ (0.1~mag deeper than the LAMOST designed) in two $20~deg^2$ FOVs in the SGC. The main purposes of the project are,
\begin{itemize}
  \item The completeness of LEGAS in the SGC: a simple definition of completeness of a spectroscopic survey is the ratio of valid targets to the observed targets in a certain magnitude range. Due to several reasons (e.g., fiber positioning/efficiency/collision, seeing, weather, color of targets), the spectra of some targets may not be obtained or the obtained spectra may not be able to pass the quality control, which influence the completeness. Moreover, a typical spectroscopic survey wouldn't be able to observe all targets in a single field due to the overdensity of targets and the time schedule. Thus, a straightforward solution is to observe all targets in several typical fields with repeating observations and directly address the question about completeness.
  \item The deficiencies of source selection methods: the aimed targets are usually selected based on the photometric parameters like color (stars and quasars) or morphology (galaxies). Each selection method has its own limitation. Thus, the complete survey of all targets in a typical field may help to build up the selection function and correct the luminosity function.
  \item The basic performance parameters of the LAMOST telescope. 
\end{itemize} 

Meanwhile, in addition to the main scientific goals, some other topics can also be addressed by taking advantage of the large FOV and numerous fibers of LAMOST, as well as the repeating observational strategy. For example, the physical properties of galaxies, identification and kinematic analysis of galaxy clusters, spectroscopic time-series analysis of variable sources (AGNs and variable stars; \citealt{Cao2016}), identification and analysis of luminous infrared galaxies and infrared excess stars \citep{Lam2015, Wu2016}. The project has considerable scientific value since it is the most complete dataset in LEGAS up to now. 

In this paper, we mainly focus on the description of the LaCoSSPAr project and the derived spectroscopic redshift catalog. The sample selection and observation are presented in \textsection 2. The data reduction and spectroscopic redshift catalog are described in \textsection 3 and \textsection 4, respectively. \textsection 5 depicts the general properties of sample galaxies. The summary is given in \textsection 6. Throughout this paper, we assume the cosmological parameters as $\Omega_{m}=0.3$, $\Omega_{\Lambda}=0.7$, $H_{0}=70~km~s^{-1}~Mpc^{-1}$.

\section{Sample Selection and Observation}

\begin{figure}
\begin{center}
\includegraphics[clip, trim=1cm 12.5cm 0.5cm 3cm, scale=0.45]{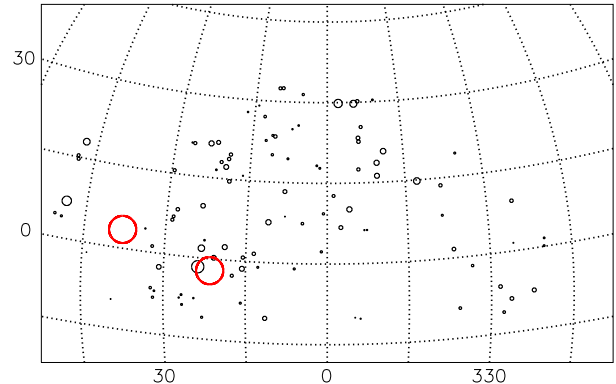}
\caption{A schematic diagram of LaCoSSPAr FOVs (red circles) in the region of South Galactic Cap u-band Sky Survey (SCUSS). The distribution of galaxy clusters from Abell rich galaxy clusters catalog are shown as circles with different sizes indicating the radius of each cluster. The x and y axes are the right ascension (R.A.) and declination (Dec.), respectively.\label{abell}}
\end{center}
\end{figure}

The LaCoSSPAr project is part of the LEGAS that mainly focus on the completeness of the galaxy survey. Based on the purpose, the restriction of the telescope guide star (there must be a central guide star with $V<8~mag$ and at least three out of four auxiliary guide CCDs obtain available guide stars with $V<17~mag$; enough number of guide stars are used to make accurate pointing of telescope and positioning of fibers), the observational timescale and season (long-exposures during five nights before and after the new moon; best site condition from October to next March), we have selected two FOVs according to the distribution of galaxy clusters from Abell rich galaxy clusters catalog \citep{Abell1989} in the region of South Galactic Cap u-band Sky Survey (SCUSS, \citealt{Zhou2016}). The main scientific goals of SCUSS are to classify different types of targets in the SGC, provide input catalog and calibrate spectra for LAMOST by using u-band deep photometry (about 1.5 mag deeper than SDSS). Thus, it is reasonable to choose the FOV from SCUSS. The central coordinates of the two FOVs we selected are $R.A.=37.88^\circ$, $Dec.=3.44^\circ$ (hereafter Field A) and $R.A.=21.53^\circ$, $Dec.=-2.20^\circ$ (hereafter Field B) as shown in Figure~\ref{abell}, which represent the lower and higher density of galaxies, respectively. 

\begin{figure*}
\includegraphics[clip, trim=4.5cm 12.5cm 4.5cm 3cm, scale=0.7]{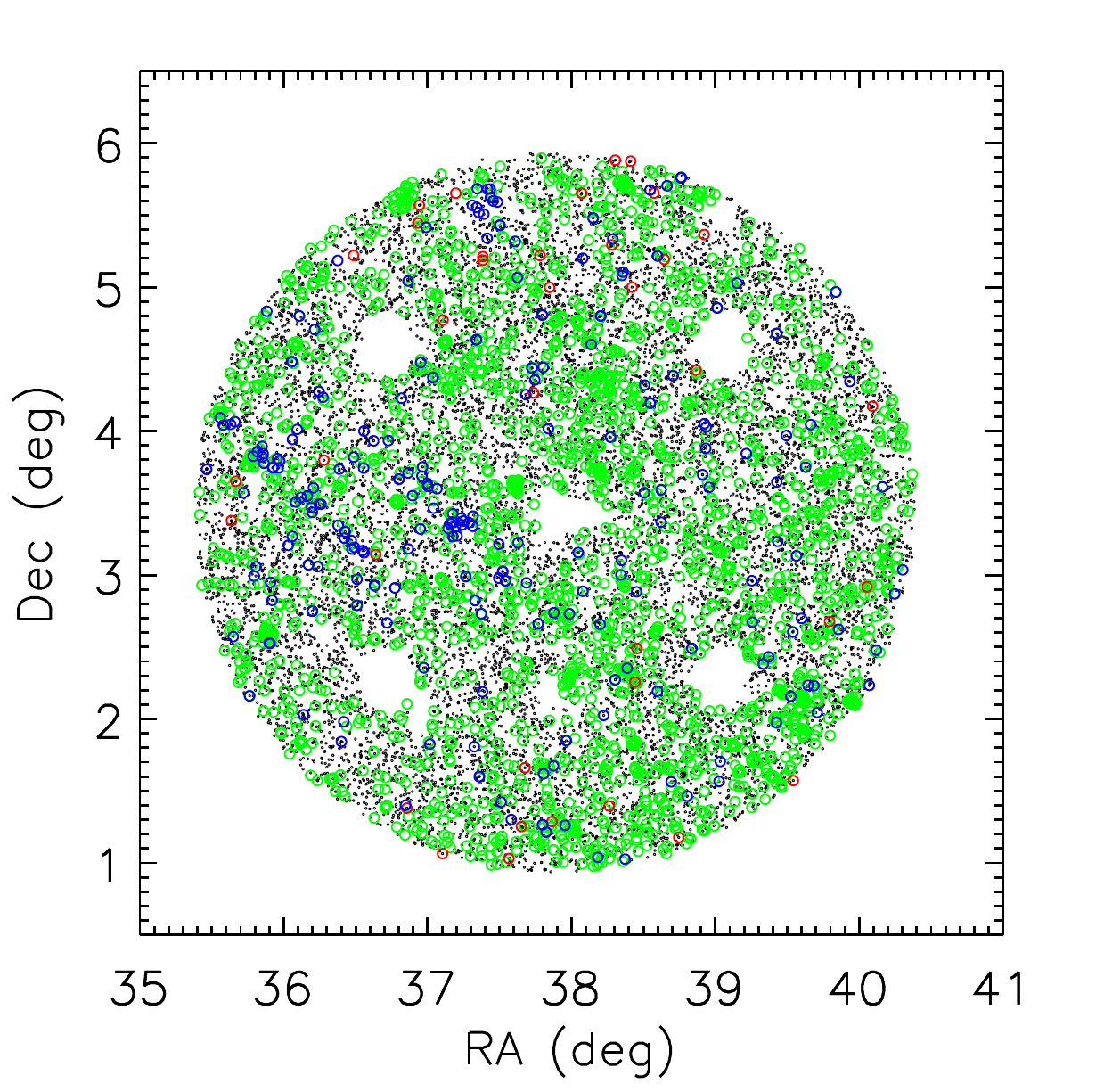}
\includegraphics[clip, trim=4.cm 12.5cm 4.5cm 3cm, scale=0.7]{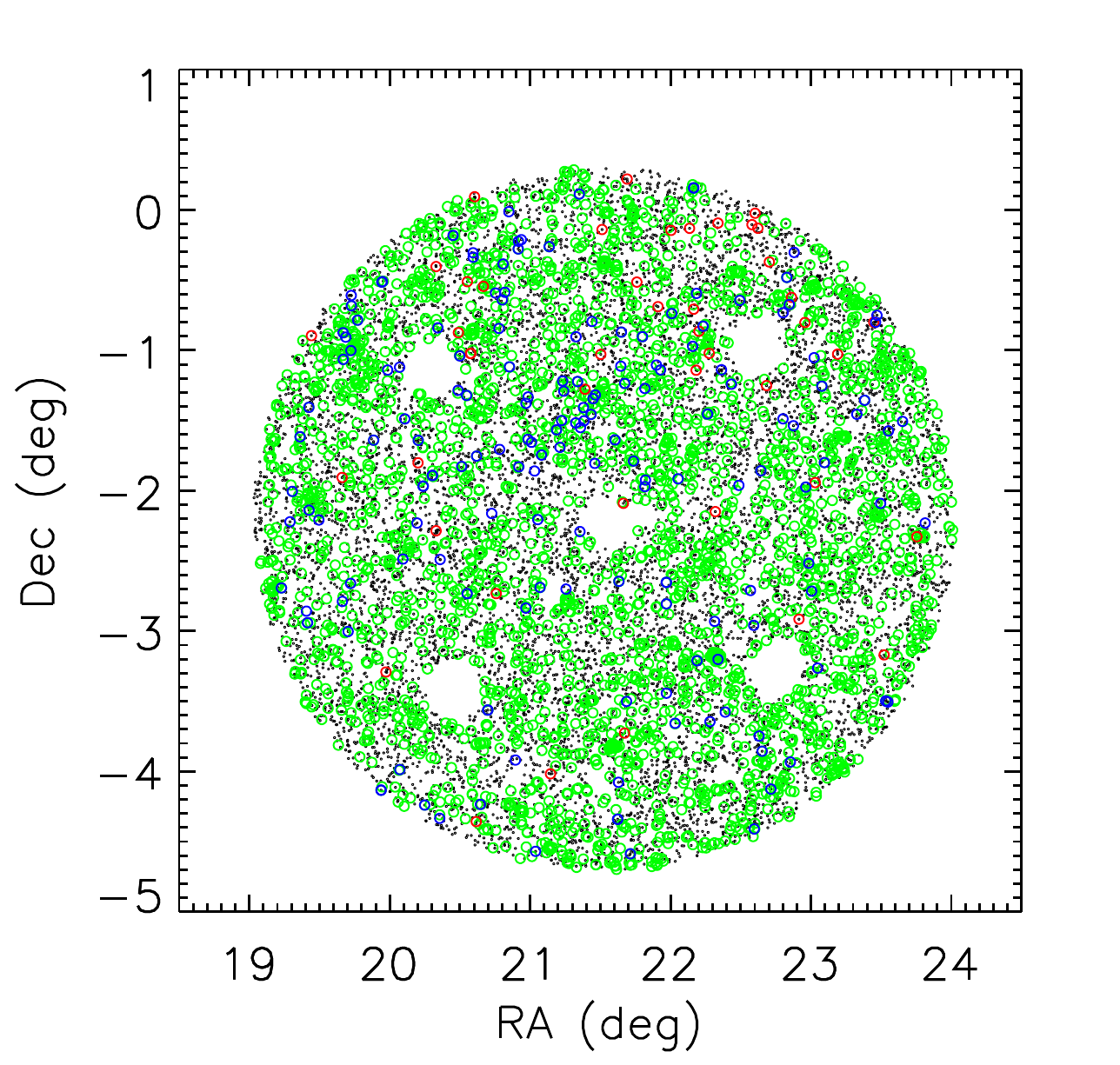}
\caption{The spatial distributions of all main targets of stars (black), galaxies (green), u-band variables (blue) and quasars (red) in Field A (left) and B (right), respectively. Five guide CCDs are indicated by the five major voids. Other small cavities are caused by the saturated stars. \label{alltarget}}
\end{figure*}

The targets in the FOV are mainly consisted of stars, galaxies, quasars, u-band variables and complementary HII regions. For the investigation of the completeness of LEGAS, the limiting magnitude of LaCoSSPAr is 0.1 mag deeper ($r=18.1~mag$) than the designed limiting magnitude of LAMOST ($r=18~mag$). The targets marked as stars and galaxies are selected from SDSS Data Release 9 (DR9) PhotoPrimary database, which contains the best version of each object along with the classification of `Star'/`Galaxy', with r-band PSF-magnitude ($r_{psf}$)/Petrosian-magnitude ($r_{petro}$) range from 14.0 to 18.1 mag \citep{Ahn2012}. The quasars candidates are collected based on the X-ray sources from XMM-Newton/Chandra/ROSAT, radio sources from FIRST/NVSS and strict color-color criteria of optical-infrared data from SDSS/UKIDSS/WISE with the same limiting magnitude (see \citealt{Ai2016} and also \citealt{Wu2010, Wu2012} for details). The u-band variables are chosen by comparing the u-band PSF-magnitude between SDSS and SCUSS with difference larger than 0.2 mag and a limiting magnitude down to $u=19.0~mag$ (see \citealt{Cao2016} for details). The complementary HII regions, which are used to fill the vacant fibers in the repeated observations (describe below), are selected based on the visual inspection of SCUSS images of the nearby galaxies. The spatial distributions of all main targets in the FOVs are shown in the Figure~\ref{alltarget}. 

Given the consideration of both faint (to obtain enough SNR) and bright end (to avoid saturation and cross-talk between neighboring fibers) of the magnitude, as well as the observational efficiency (to take less observation time as much as possible due to the tight time schedule of LAMOST scientific survey), all targets in the FOV have been divided into two kinds of plates, Bright (B-plate, $r=14.0\sim16.0~mag$) and Faint (F-plate, $r=16.0\sim18.1~mag$), according to their r-band magnitude. The typical exposure time for B- and F-plate are $3\times600~s$ and $3\times1800~s$, respectively. Due to the weather condition, time schedule of daily observation and evaluation of telescope performance, the exposure time varies for some plates. 

The observation of LAMOST is automatically managed by the Survey Strategy System (SSS), which aims at making LAMOST an efficient project by determining the observational schedule, setting up the input catalog and corresponding fiber positions, arranging the observational process \citep{Yuan2008}. Apart from the automatic process, manual intervention also has been performed to ensure that the observations of LaCoSSPAr are conducted during the dark and gray nights. As we mentioned before, the main purposes of our project are evaluation of the completeness of LEGAS and the deficiencies of source selection methods, which require the complete observation of \textit{ALL} sources in the FOV as much as possible, especially for the galaxies. Given the high number density of stars ($>11000$/FOV) and clustering of galaxies, the observation has to be repeated to achieve the goals, which also provides us an opportunity to observe variable sources with different time intervals. Thus, the highest priority (higher priority, larger chance to be observed) has been assigned to the u-band variables and quasars to keep them in each observation as far as possible. Galaxies have been given the secondary priority. Meanwhile, as the observation continues, there are less galaxies need to be observed and more vacant fibers. Thus, the low SNR galaxies ($SNR_{r}<3$) in previous observations and HII regions also have been given the same priority as galaxies to fill the extra vacant fibers. Stars are set to be the lowest priority. At the same time, we have also applied multiple observations for B-plate to try to understand the binary rate in certain magnitude range, since the exposure time for the B-plate is much less than the F-plate.

We conduct the observation from September 2012 to January 2014 (except July and August which are the maintain period of the telescope). Field A contains 2519 galaxies and 13930 stars in which 2447 ($\sim97\%$) and 13278 ($\sim95\%$) of them have been observed with 5 B- and 12 F-plates (redundant F-plates are caused by a technical issue), respectively. Field B contains 3104 galaxies and 11381 stars in which 2995 ($\sim97\%$) and 10224 ($\sim90\%$) of them have been observed with 6 B- and 5 F-plates, respectively. Due to several reasons (e.g., the status of the instruments, seeing, spectral calibration process and so on), not all the observed spectra are available. Moreover, since December 2012, the targets in the range of 10" from the bright stars (R1- or R2-band magnitude less than 12 in USNOB1 catalog) have been masked to prevent the contamination. The date, plate name, exposure time and dome seeing are listed in Table~\ref{tbl1}.

\section{Data Reduction}

The raw data have been reduced with LAMOST 2D and 1D pipelines which include bias subtraction, flat-fielding through twilight exposures, cosmic-ray removal, spectrum extraction, wavelength calibration, sky subtraction, flux calibration and exposure coaddition (see \citealt{Luo2004, Luo2012, Luo2015} for details). Due to several problems of the spectrum (e.g., low SNR, remaining sky background residuals, combining the blue and red parts of the spectrum, uncertainty of spectral calibration), the number of redshifts measured by the pipeline is limited (about 1/3 of the total observed galaxies) and no physical parameters are derived. Therefore, we have to adopt additional processes to deal with LAMOST 1D spectrum which makes the measurement of redshift easier by visually inspecting of emission/absorption lines. 

Briefly speaking, since the sky subtraction is a tricky and complicated step in the spectral reduction \citep{Bai2017}, it is possible that some artificial emission/absorption lines are actually caused by the sky subtraction (underestimation/overestimation of the sky background) rather than the real physical mechanism. Especially, for our case, the contamination is worse as lower SNR obtained.

\begin{figure}
\begin{center}
\includegraphics[clip, trim=1.5cm 12.5cm 3cm 3cm, scale=0.5]{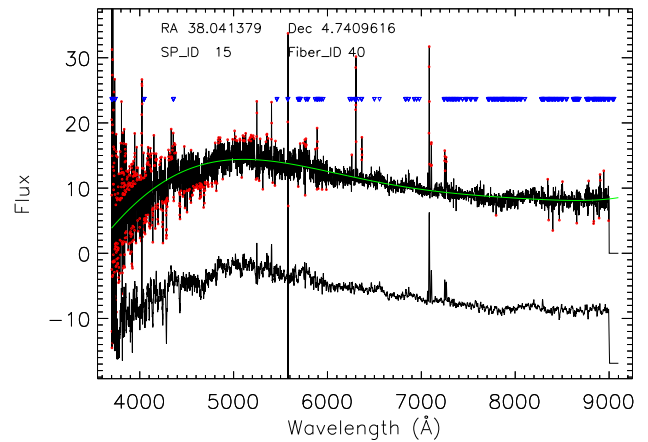}
\caption{An example of post-processing of emission line galaxy (the same as for absorption line galaxy and quasar). From top to bottom: the sky lines (blue open triangles), original spectrum, reduced spectrum. The robust polynomial fitting and 3$\sigma$-points on the original spectrum are shown as green curve and red points, respectively (see text for details). Coordinate, spectrograph ID and fiber ID are marked on the top.\label{sp_reduction}}
\end{center}
\end{figure}

Based on this assumption, each sky line is marked on the original spectrum at the beginning. The continuum has been fitted by a six-order robust polynomial and subtracted from the original spectrum. The robust sigma ($\sigma$) of the residual has been calculated and points with absolute value larger than 3$\sigma$ are marked. If the 3$\sigma$-points are overlapped with the sky lines, then they are probably caused by the sky subtraction rather than the true emission/absorption. Thus, those 3$\sigma$-points are replaced by the continuum fitting. It is worth to notice that there is still a small chance (less than 1\% according to our inspection) that real emission/absorption lines are exactly superposing on the sky lines. For such case, the real emission/absorption lines may be also replaced by our process. However, since the measurement of redshift is based on cross-matching of multiple lines, this problem can be overcame. For broad emission lines (e.g., AGNs and quasars), the overall line profile wouldn't be changed by our process. Finally, we conduct the boxcar-smoothing through the entire spectrum with a window of 5 points. The examples of original spectrum and reduced spectrum are shown in Figure~\ref{sp_reduction}. With the additional post-processing, the true emission/absorption lines become more clear and distinct, especially for low SNR galaxies. 

\begin{figure*}
\begin{center}
\includegraphics[clip, trim=2cm 14.5cm 1cm 3cm]{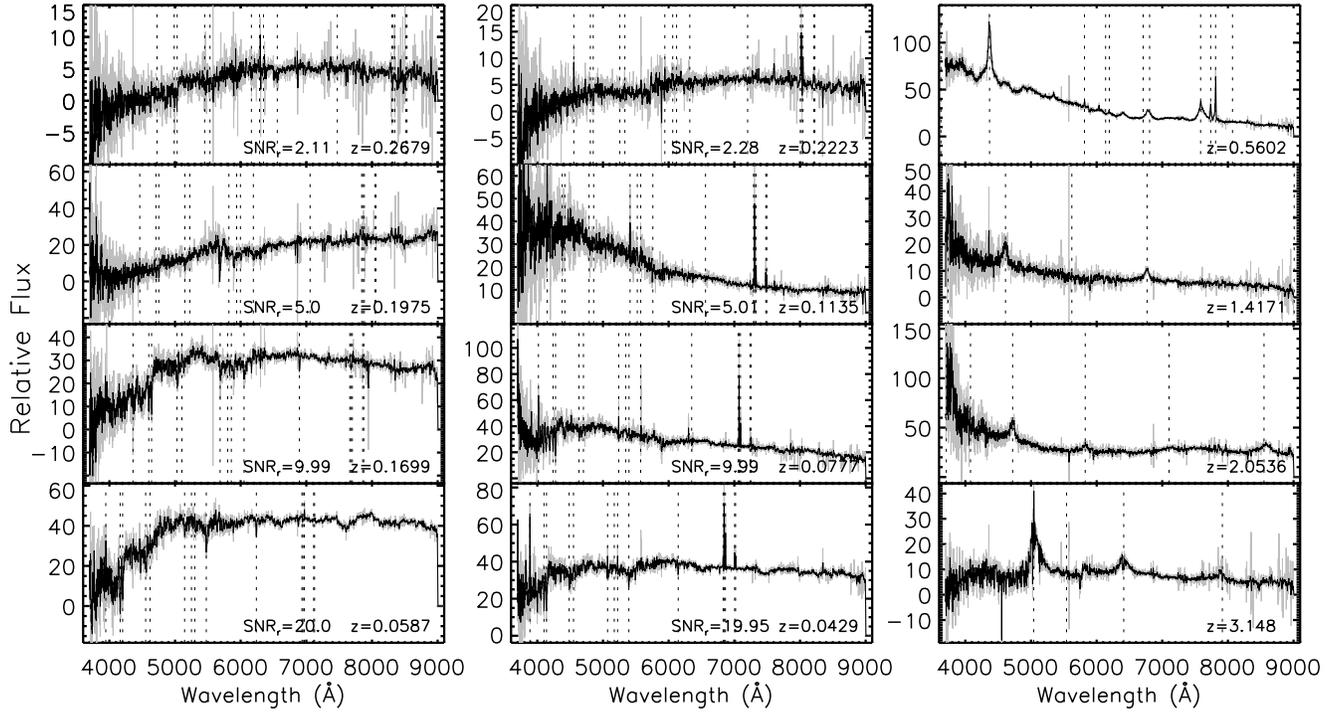}
\caption{Examples of absorption line galaxies (left), emission line galaxies (middle) and quasars (right). The $SNR_r$ (only for absorption and emission line galaxies) and redshift are marked in the bottom right of each diagram. \label{sp_example}}
\end{center}
\end{figure*}

The next step is to measure the redshift. In total, there are more than 10,000 spectra have been visually inspected. Each spectrum of galaxy has been double checked by at least two individuals. All galaxies have been divided into two categories, emission and absorption line galaxies, depending on whether the obvious emission lines appeared or not. The measurement of redshift starts with matching the emission lines, then the absorption lines. For emission lines, the [O\uppercase\expandafter{\romannumeral2}] $\lambda$3727, H$\beta$ $\lambda$4861, [O\uppercase\expandafter{\romannumeral3}] $\lambda\lambda$4959,5007, H$\alpha$ $\lambda$6563, [N\uppercase\expandafter{\romannumeral2}] $\lambda\lambda$6548,6583, [S\uppercase\expandafter{\romannumeral2}] $\lambda\lambda$6717,6731 have been adopted. For absorption lines, the Ca\uppercase\expandafter{\romannumeral2} HK $\lambda\lambda$3933,3969, G-band $\lambda$4300, Mg\uppercase\expandafter{\romannumeral1} $\lambda$5172, NaD $\lambda$5893, H$\alpha$ $\lambda$6563 along with the obvious feature of D4000 break have been adopted. For quasar, the additional emission lines of C\uppercase\expandafter{\romannumeral4} $\lambda$1549, C\uppercase\expandafter{\romannumeral3}] $\lambda$1909, C\uppercase\expandafter{\romannumeral2}] $\lambda$2326 and Mg\uppercase\expandafter{\romannumeral2} $\lambda$2798 have been considered. The perplexing emission/absorption lines caused by the remaining sky background residuals, such as $\lambda$4360, 5577, 6300, 6830, 7606, \textgreater8000$\textup{\AA}$, and the overlapping region of the blue and red spectrograph arms, e.g., $5700\sim5900\textup{\AA}$, are also appearing on the spectrum. These confused lines have been carefully examined to avoid wrong measurement. For each normal galaxy, the redshift has been reliably confirmed with at least four emission/absorption features, while three features are applied for quasars. The uncertainty of redshift by visual inspection is about $\sigma_{z}/(1+z)<0.001$. Figure~\ref{sp_example} shows the typical examples of absorption line galaxies, emission line galaxies and quasars with r-band SNR ($SNR_r$; only for absorption and emission line galaxies) and redshifts labeled.

During the inspection, we notice that a small portion of the absorption line galaxies show a similar pattern with H$\alpha$ absorption, weak [N\uppercase\expandafter{\romannumeral2}] $\lambda$6583 and [S\uppercase\expandafter{\romannumeral2}] $\lambda\lambda$6717,6731 emissions, which may indicate the overlapped H$\alpha$ emission and continuum absorption, or the obscured AGN embedded in the center of galaxy. However, we cannot confirm it without further high resolution/SNR observation. Thus, those galaxies are still categorized as absorption line galaxy. There are also a few instances of star-galaxy or galaxy-galaxy overlapping, in which the magnitude and color may not be reliable due to the contamination. We have visually inspected the SDSS images for all targets since some of them are HII regions belonging to the giant galaxies. Still, we keep those targets in the catalog with labels. The obvious AGNs (LINERs and Seyferts) also have been tentatively identified during the process and double checked with $W2-W3$/$W1-W2$ diagram (\textsection 5) since most of the galaxies with $[3.4]-[4.6] > 0.7$ are strong AGNs \citep{Jarrett2011, Lam2015}.

\section{The Spectroscopic Redshift Catalog}

\begin{figure*}
\begin{center}
\includegraphics[clip, trim=2cm 14.5cm 2cm 5.5cm]{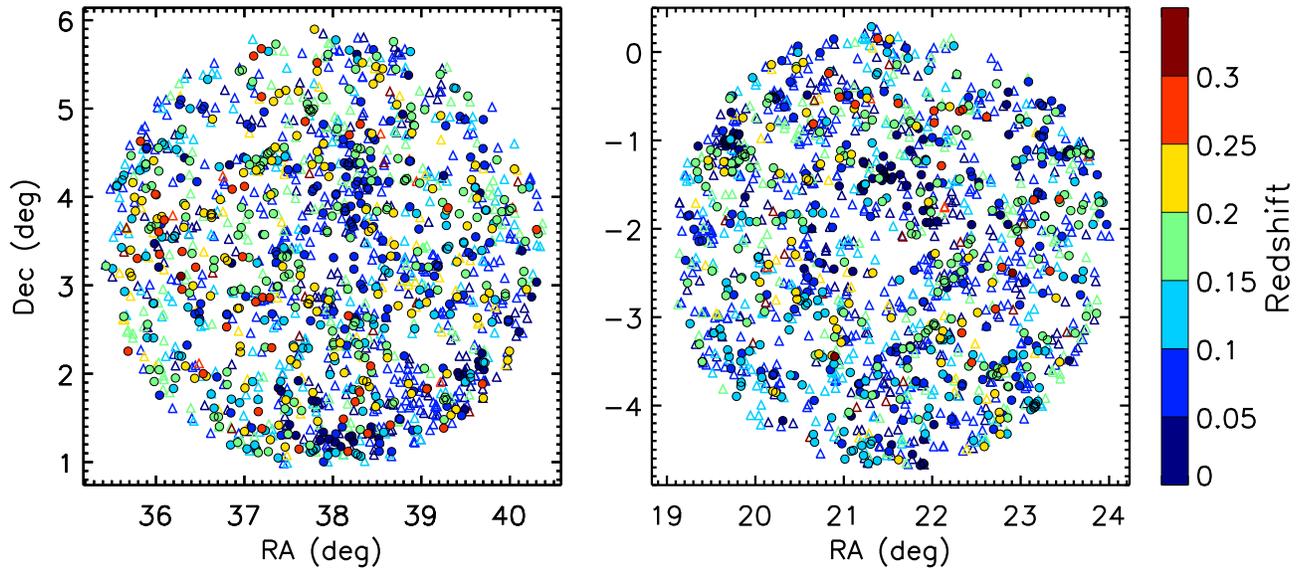}
\caption{The spatial distribution of all absorption (solid circle) and emission line (open triangle) galaxies in Field A (left) and B (right). Each galaxy has been color coded by redshift (same below). \label{all_ea_spatial_distribution}}
\end{center}
\end{figure*}

In total, there are 2447 galaxies have been observed and 1528 redshifts ($\sim62\%$ of observed galaxies; including 13 quasars) have been measured in Field A with median value of $z=0.125$ and range from 0.001 to 1.598 (99\% at $z<0.3$). Among them, there are 623 (40.8\%) absorption and 905 (59.2\%) emission line galaxies, respectively. In Field B, there are 2995 galaxies have been observed and 1570 redshifts ($\sim53\%$ of observed galaxies; including 16 quasars) have been measured with median value of $z=0.0952$ and range from 0.006 to 3.148 (99\% at $z<0.3$). Among them, there are 569 (36.2\%) absorption and 1001 (63.8\%) emission line galaxies. The spatial distributions of all absorption and emission line galaxies in both fields are shown in Figure~\ref{all_ea_spatial_distribution}. It is noticeable that even if there is a number difference of more than 500 between the observed galaxies in the two fields, yet the difference between the numbers of measured redshifts is less than 50. The main reason is that, due to the redundant observations, a fair amount of galaxies in Field A have been observed with better seeing and more accurate fiber positioning, which results in more derived redshifts.

\begin{figure*}
\begin{center}
\includegraphics[clip, trim=2cm 12.5cm 2cm 2.5cm]{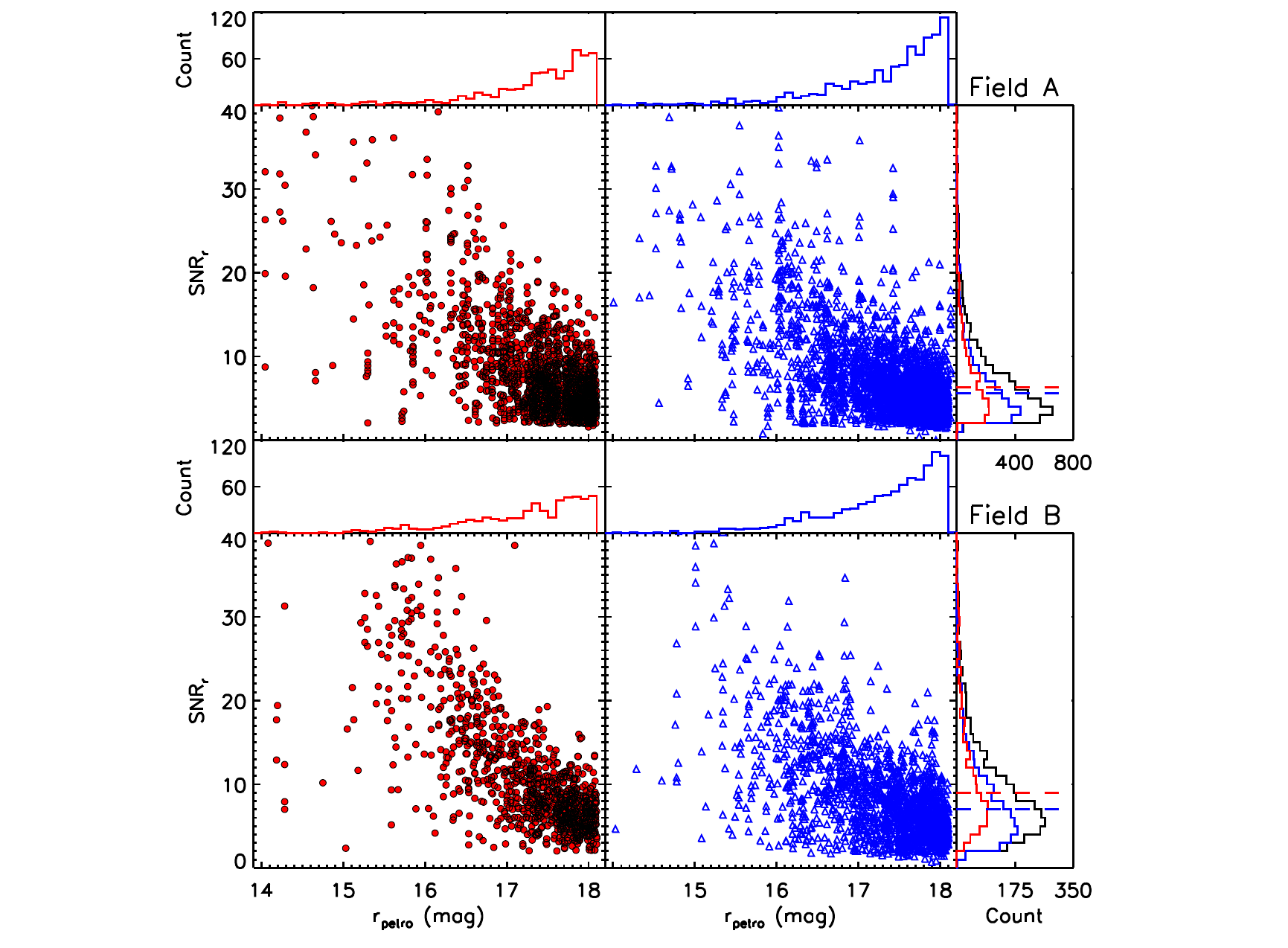}
\caption{The $r_{petro}$ versus $SNR_r$ diagrams for absorption (left column) and emission line galaxies (right column) in Field A (top row; 4525 individual observations) and B (bottom row; 2636 individual observations), respectively. The histograms of magnitude (individual galaxy; at the top of each panel) and $SNR_r$ (individual observation; on the right of each row) also have been shown on the diagram. The dashed lines indicate the median value of $SNR_r$ for absorption (red) and emission line galaxies (blue). The peaks and median values of $SNR_r$ are around 3 and 6 for Field A and 5 and 8 for Field B, respectively. The distribution of $SNR_r$ shows that it is possible to derive redshift from low SNR galaxies with our post-processing and visual inspection, since the $SNR_r$ is less than 5 for about 41\% and 24\% of galaxies in Field A and B, respectively.\label{rmag_snrr_z}}
\end{center}
\end{figure*}

ALL galaxies with measured redshift are cross-matched with LAMOST DR4 to obtain the $SNR_r$ calculated by using the inverse variance \citep{Luo2015}. Each individual observation has been taken into account instead of the average value for individual galaxy due to the changing of the SNR under different observational conditions. Figure~\ref{rmag_snrr_z} shows the distribution of $r_{petro}$ versus $SNR_r$ for all galaxies with 4525 individual observations in Field A and 2636 in Field B, respectively. The histograms of magnitude (individual galaxy) and $SNR_r$ (individual observation) also have been shown on the diagram. As the diagram shown, the peaks and median values of $SNR_r$ are around 3 and 6 in Field A and 5 and 8 in Field B, respectively. Majority of galaxies are found to have $SNR_r$ from 3 to 20. The overall distribution of SNR and magnitude/SNR in Field B is better than Field A due to the major renovation of the telescope infrastructure during the summer of 2013, which provides higher throughput and better pointing accuracy. The distribution of SNR also shows that it is possible to derive redshift from low SNR galaxies with our post-processing and visual inspection, since the $SNR_r$ is less than 5 for about 41\% and 24\% of galaxies in Field A and B, respectively.

\begin{figure*}
\center
\includegraphics[clip, trim=2cm 14.5cm 2cm 4cm]{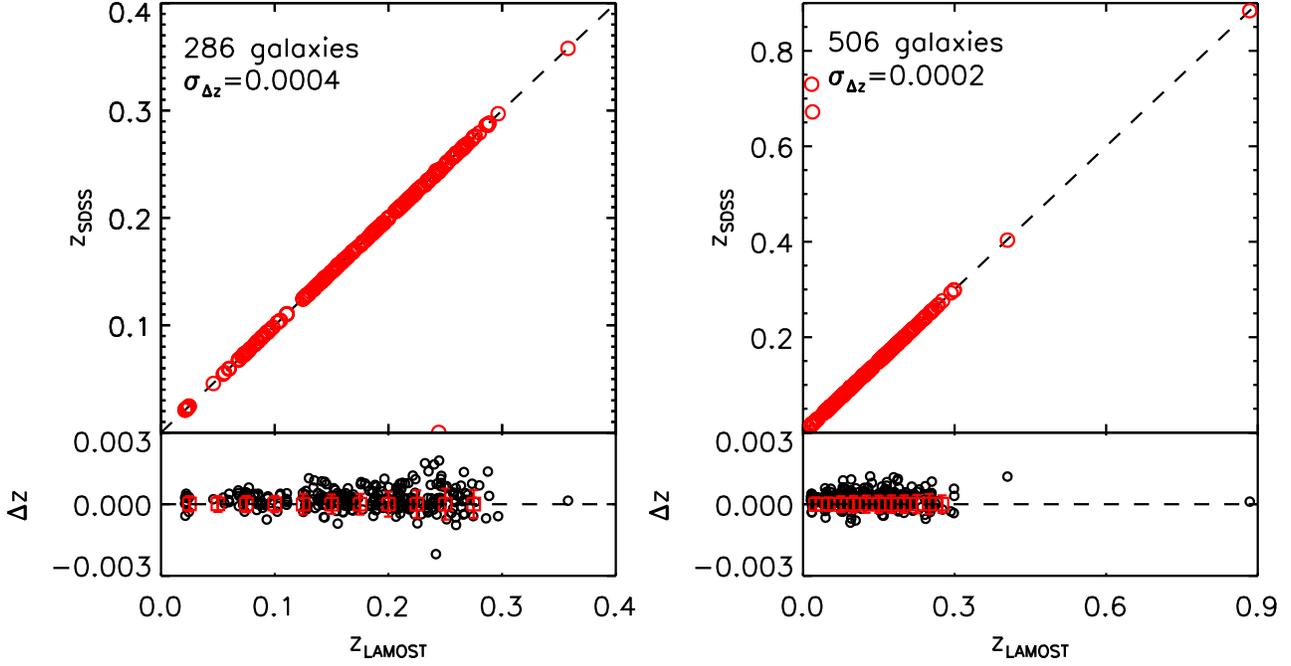}
\caption{The comparison of $z_{sp}$ between LAMOST and SDSS within a cross-matching radius of 1" for Field A (left) and B (right). All matched galaxies have almost identical redshift. One outlier in Field A ($z_{LAMOST}\sim0.24$) is due to the overlapping of an absorption line galaxy and a late-type star. The two outliers in Field B are caused by the low SNR of spectra both in LAMOST and SDSS. \label{lamost_sdss_z}}
\end{figure*}

Our result also has been compared with SDSS DR9 to demonstrate the reliability as shown in Figure~\ref{lamost_sdss_z}. In total, there are 286 and 506 galaxies which both have LAMOST and SDSS spectroscopic redshift ($z_{sp}$) within a cross-matching radius of 1" in Field A and B, respectively. All matched galaxies have almost identical redshift between LAMOST and SDSS with $\sigma_{\Delta z}=0.0004$ for Field A and $\sigma_{\Delta z}=0.0002$ for Field B. In Field A, there is only one outlier, which has been confirmed as an overlapping of an absorption line galaxy and a late-type star by SDSS image inspection. For Field B, the two outliers are both low SNR galaxies either in LAMOST or SDSS which prevents the further confirmation.

\begin{figure*}
\begin{center}
\includegraphics[clip, trim=2cm 12.5cm 2cm 2.5cm]{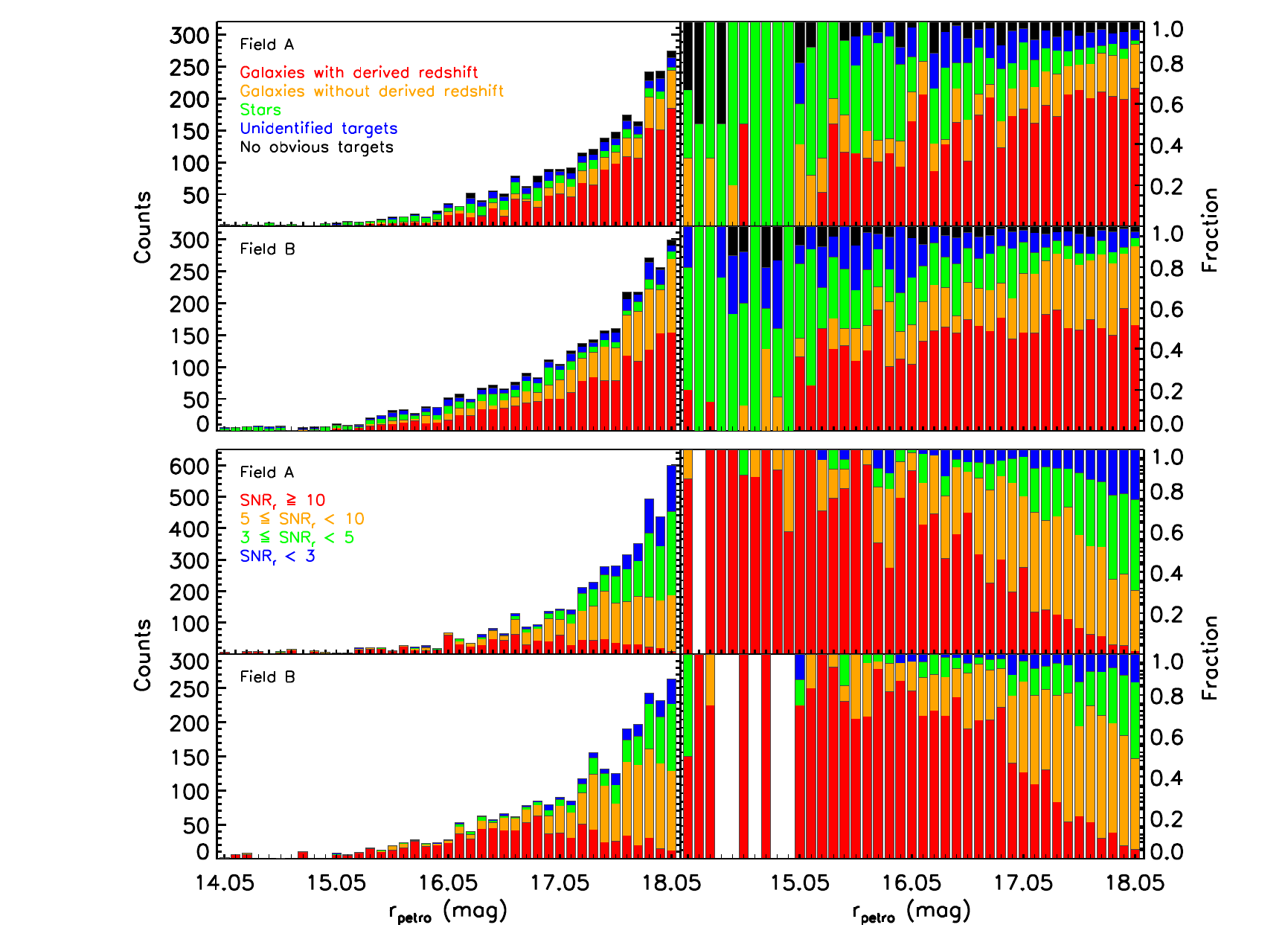}
\caption{Upper panel: the histograms of $r_{petro}$ for all input targets (visually inspected and divided into five categories as shown in the diagram) along with their fraction within each bin of 0.1 magnitude. According to the visual inspection, up to 1/4 (sum of category 3, 4 and 5) of the input targets might be unreliable. Lower panel: the histograms of $r_{petro}$ for galaxies with derived redshift. Each individual observation has been taken into account and color coded by $SNR_{r}$. In given magnitude, the SNR of Field B is better than Field A ($\sim10\%$ in average for targets with $SNR_{r}\ge5$) due to the major renovation of the telescope infrastructure. \label{all_category}}
\end{center}
\end{figure*}

To understand better the completeness and selection effect of LEGAS and constrain the broader survey, we have also checked the morphologies of all input targets by visually inspecting the SDSS images. All targets in the input catalog have been divided into five categories: galaxies with derived redshift, galaxies without derived redshift, stars, unidentified targets (too fuzzy) and no obvious targets. The upper panel of Figure~\ref{all_category} shows the histograms of $r_{petro}$ for all five categories along with their fraction within each bin of 0.1 magnitude. The percentage for each category is about 59\%, 16\%, 12\%, 8\%, 5\% in Field A and 50\%, 26\%, 12\%, 9\%, 3\% in Field B, respectively. In principle, the result may indicate that up to 1/4 (sum of category 3, 4 and 5) of the input targets for a typical extra-galactic spectroscopic survey might be unreliable based on the visual inspection. The improvement of classification algorithm of the prior photometric survey could significantly improve the efficiency of the following spectroscopic survey. This may be important for the forthcoming large-scale spectroscopic surveys like 4MOST, MOONS, WEAVE, DESI, PFS, MSE and so on. The part of galaxies without derived redshift in our sample is mainly due to the clustering of galaxies, unavailable spectra, low SNR, uncertainties of spectral calibration and offset of the fiber positioning. However, we also notice that the situation can be improved that additional $\sim10\%$ of redshifts are obtained by the redundant observations. Although this improvement is inaccessible in the normal survey, it might be achieved in some small-scale deep observations. The lower panel of Figure~\ref{all_category} shows the histogram of $r_{petro}$ for galaxies with derived redshift. Each individual observation has been taken into account and color coded by $SNR_{r}$. As described above, it can be seen that, in given magnitude, the SNR in Field B is better than Field A ($\sim10\%$ in average for targets with $SNR_{r}\ge5$) due to the major renovation of the telescope infrastructure. The results derived from Field A have been reported to the Working Group of LEGAS and LAMOST Scientific Committee after the first year of survey. Combined with other scientific and technical reports, it results in a major change of survey strategy that, more observation time has been allocated to the LAMOST Experiment for Galactic Understanding and Exploration (LEGUE). Meanwhile, the deep survey ($r=16\sim18.8~mag$ for normal and blue galaxies; \citealt{Luo2015}) of LEGAS is mainly focusing on two stripes ($|b|>30$, $Dec.\approx5^\circ$ in the SGC and $Dec.\approx30^\circ$ in NGC with a width of $\sim10^\circ$) and the shallow survey ($r=14-16.8~mag$) covering both NGC and SGC remains unchanged.

To obtain more information about each target, we also cross-match our sample with GALEX and ALLWISE catalog \citep{Martin2005, Wright2010}. For GALEX, a search radius of 4" has been applied to cross-match with the original SDSS coordinate and the closest match detected either in far-ultraviolet (FUV) or near-ultraviolet (NUV) band has been selected, which result in 1031 (67.5\%) and 1272 (81.0\%) matches in Field A and B, respectively. The significant difference ($\sim15\%$) in the UV detection between the two fields is due to the shallower depth of GALEX survey in Field A. For ALLWISE, a search radius of 6" has been adopted which lead to 1511 (98.9\%) and 1544 (98.3\%) matches in W1 band. Meanwhile, since the ALLWISE catalog has already been cross-matched with 2MASS, there are also 1265 (82.8\%) and 1295 (82.5\%) matches in J band. Combined with original SDSS photometric data, it forms the multi-wavelength catalog from FUV to mid-infrared (MIR). 

Table~\ref{tbl2} and ~\ref{tbl3} contain the full information of coordinates (R.A. and Dec.), galaxy types (absorption/emission line galaxy), redshifts and multi-wavelength data for Field A and B, respectively. For each table, the column 1 is the index of each target. The columns 2 and 3 are the coordinate. The column 4 is the galaxy type and column 5 is the redshift. The columns 6 to 37 are the multi-wavelength data from FUV to 22\micron~(W4) with corresponding errors and SNR (only for the ALLWISE data). The columns 38 to 44 are the k-corrected FUV to z-band magnitudes (see below). The column 45 is the comments about special targets (e.g. HII regions, star-galaxy/galaxy-galaxy overlapping, AGNs and so on). Targets without detection, or the errors of the upper limits ($SNR<3$; only for the ALLWISE data) in each passband are assigned with -999.0.

\section{The General Properties of the Sample Galaxies}

Due to several problems of the spectrum and lack of derived physical parameters from the pipeline as described before, the multi-wavelength data have been used to evaluate the general properties of our sample. 

\begin{figure*}
\begin{center}
\includegraphics[clip, trim=2cm 14.5cm 1.5cm 4cm]{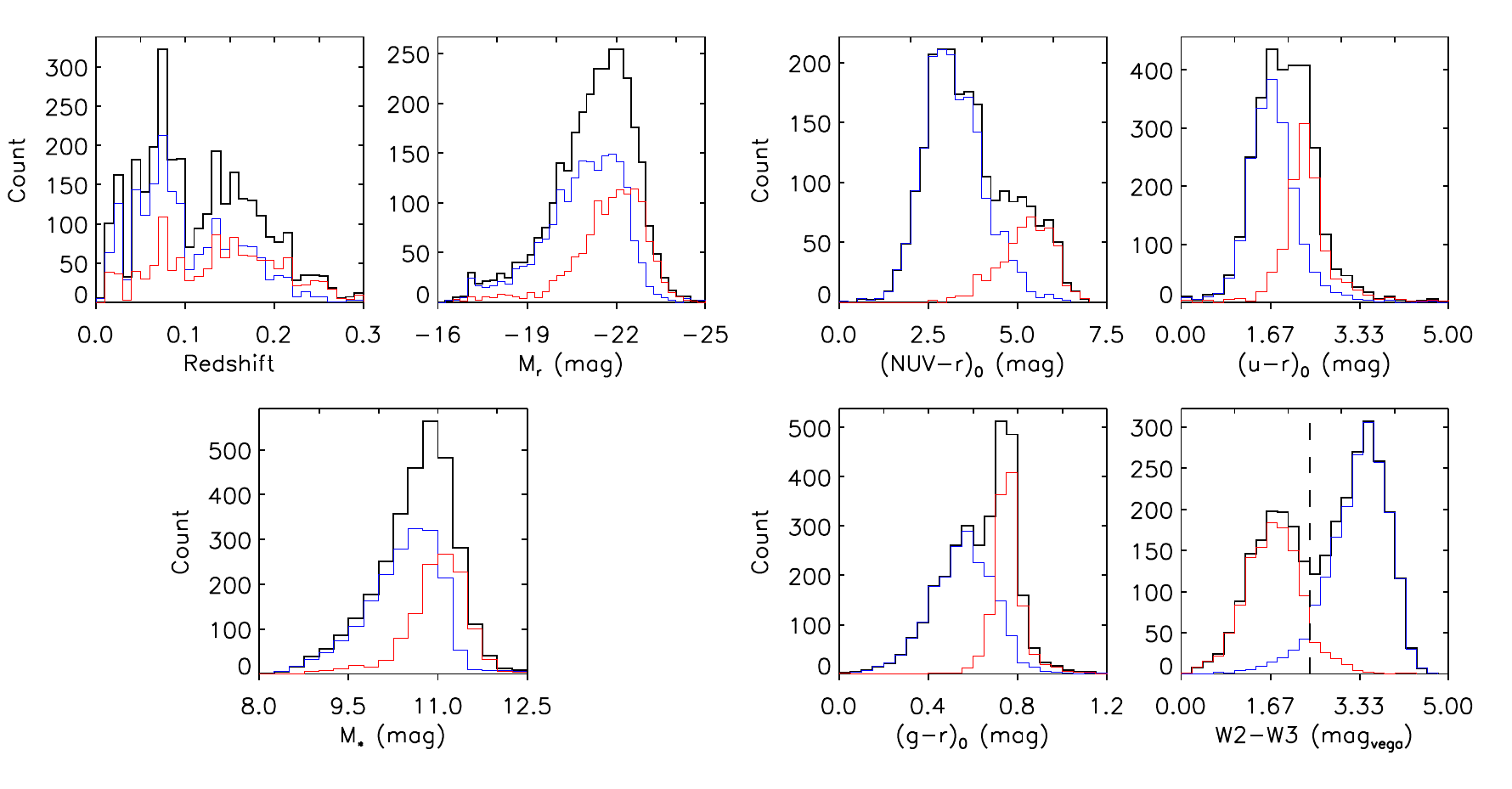}
\caption{The histograms of redshift, $M_r$, $M_*$, rest-frame colors of $(NUV-r)_0$, $(u-r)_0$, $(g-r)_0$ and observed color of $W2-W3$ for the entire sample (black outline). For convenience, we confine the range of x-axis as shown in each panel. The red and blue color indicate the absorption and emission line galaxies, respectively. From the histogram of $W2-W3$ color, it has been revealed that the MIR-detected absorption (91.3\%) and emission line galaxies (93.3\%) can be well separated by an empirical criterion of $W2-W3=2.4$. \label{all_histogram}}
\end{center}
\end{figure*}

We use the algorithm from \citet{Chilingarian2012} to calculate the k-correction for the ultraviolet data (FUV and NUV; \citealt{Taylor2005}). The $kcorrect~v4\_2$ has been used to compute the k-correction for SDSS data \citep{Blanton2007}. No k-correction has been applied to the infrared data. The k-corrected GALEX and SDSS data are given in the AB system (the Galactic extinction corrected SDSS magnitudes are converted to the corresponding magnitudes in the AB system\footnote{http://www.sdss3.org/dr9/algorithms/fluxcal.php\#SDSStoAB}), while the 2MASS and WISE data are given in the Vega system. The $M_*$ has been calculated by using r-band absolute magnitude ($M_r$) and rest-frame $g-r$ color ($(g-r)_0$) \citep{Bell2003, Bernardi2010} as:
\begin{equation}
log_{10}\left(\frac{M_{*}}{M_{\sun}}\right)=1.097(g-r)_0-0.406-0.4(M_{r}-4.67)-0.19z,
\end{equation}
where $z$ is the redshift of each target. The histograms of redshift, $M_r$, $M_*$, rest-frame colors of $(NUV-r)_0$, $(u-r)_0$, $(g-r)_0$ and observed color of $W2-W3$ for the entire sample are shown in Figure~\ref{all_histogram}. The red and blue colors indicate the absorption and emission line galaxies, respectively (same below).

\begin{figure}
\begin{center}
\center
\includegraphics[clip, trim=2cm 0cm 2cm 0cm, scale=0.7]{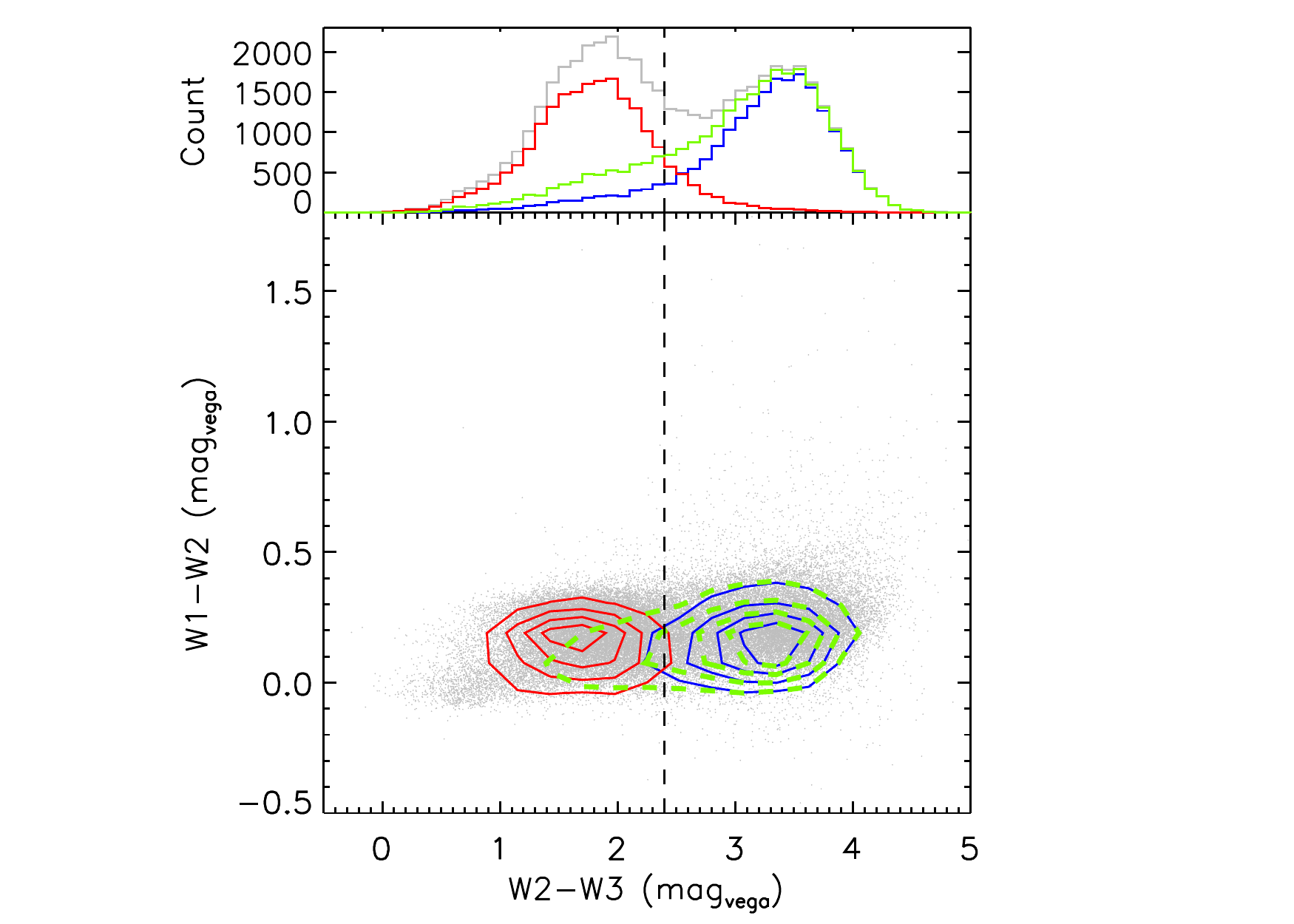}
\caption{The $W2-W3$/$W1-W2$ diagram of the control sample galaxies from MPA/JHU value-added galaxy catalog along with the histogram of $W2-W3$ color. Red and blue contours indicate the absorption (`unclassifiable') and emission line galaxies (`SF', `low SNR SF', `composite'), respectively. Additional green dashed contour represents the inclusion of `non-LINER AGNs' and `low SNR LINER' in the sample of emission line galaxies, which heavily contaminates the boundary between absorption and emission line galaxies. \label{sdss_wise}}
\end{center}
\end{figure}

From the diagrams, it has been revealed that the MIR-detected absorption (91.3\%) and emission line galaxies (93.3\%) can be well separated by an empirical criterion of $W2-W3=2.4$ \citep{Alatalo2014}. The obvious reason for this scenario is the different dust content between these two types of galaxies that, the majority of dust-deficient, low-SF, absorption line galaxies and dust-rich, high-SF, emission line galaxies can be distinguished by the $W2-W3$ colors \citep{Li2007, Mentuch2010, Donoso2012, Meidt2012, Jarrett2017}. Here we do not set the SNR cutoff for W2 and W3 bands since it will bias the result towards the more dusty galaxies, especially for absorption line galaxies with low dust content. For further confirmation, the data from MPA/JHU value-added galaxy catalog for SDSS DR7 has been checked as the control sample \citep{Tremonti2004}. For simplicity, only the galaxies belong to the three stripes in the SGC have been selected and cross-matched with ALLWISE catalog using the same criterion as we used. All galaxies in the MPA/JHU catalog have been classified into seven categories depending on their emission line properties \citep{Brinchmann2004}. Figure~\ref{sdss_wise} shows the $W2-W3$/$W1-W2$ diagram for the control sample overlapped with the contours of absorption (red; category of `unclassifiable') and emission line galaxies (blue; categories of `SF', `low SNR SF' and `composite') along with the histogram of $W2-W3$ color. The clear separation between these two types of galaxies (88.1\% for absorption and 87.8\% for emission line galaxies) is again obvious. Further investigation shows that the inclusion of `non-LINER AGNs' and `low SNR LINER' in the sample of emission line galaxies will contaminate the separation by showing as the green dashed contour.

\begin{figure*}
\begin{center}
\includegraphics[clip, trim=2cm 14.5cm 0cm 4cm]{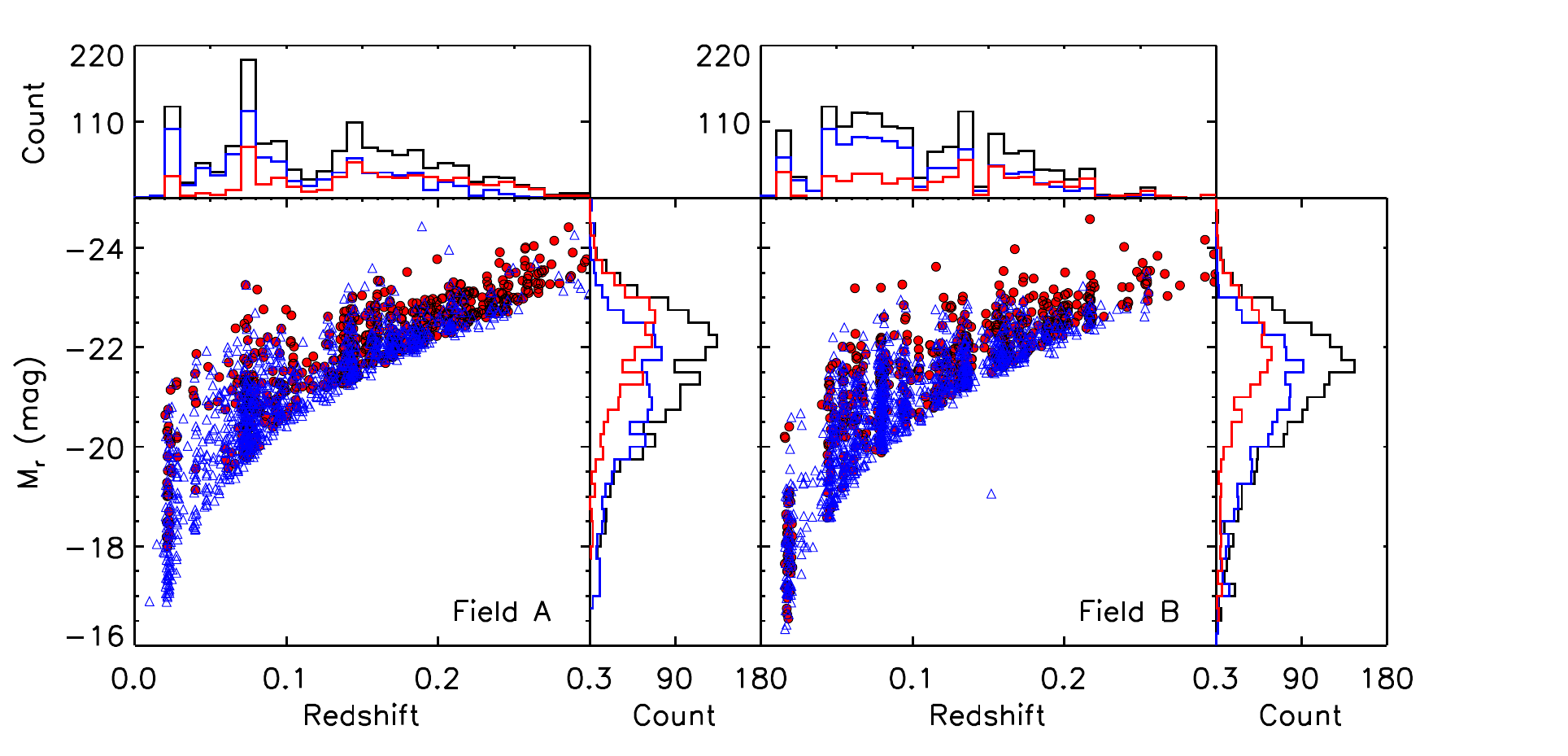}
\caption{The redshift versus $M_r$ diagrams for Field A (left) and B (right). The red solid circles and blue open triangles indicate the absorption and emission line galaxies, respectively. One outlier with $z\approx0.15$ and $M_{r}\approx-19$ in Field B is the fainter one of the two overlapped galaxies seen on the SDSS image, for which the magnitude may not be reliable due to the contamination. \label{all_z_mr}}
\end{center}
\end{figure*}

Figure~\ref{all_z_mr} shows the redshift versus $M_r$ diagram for each field along with the histograms of redshift and $M_r$. One outlier with $z\approx0.15$ and $M_{r}\approx-19$ in Field B is the fainter one of the two overlapped galaxies seen on the SDSS image, for which the magnitude may not be reliable due to the contamination. Unfortunately, at this stage, we may not be able to further quantify the completeness of our sample from this diagram, since the results are largely based on the visual inspection and some targets are missing due to the pipeline reduction/bright star masking as described in the last paragraph of \textsection 2.

\begin{figure*}
\begin{center}
\includegraphics[clip, trim=2cm 12.5cm 2cm 2cm, scale=0.9]{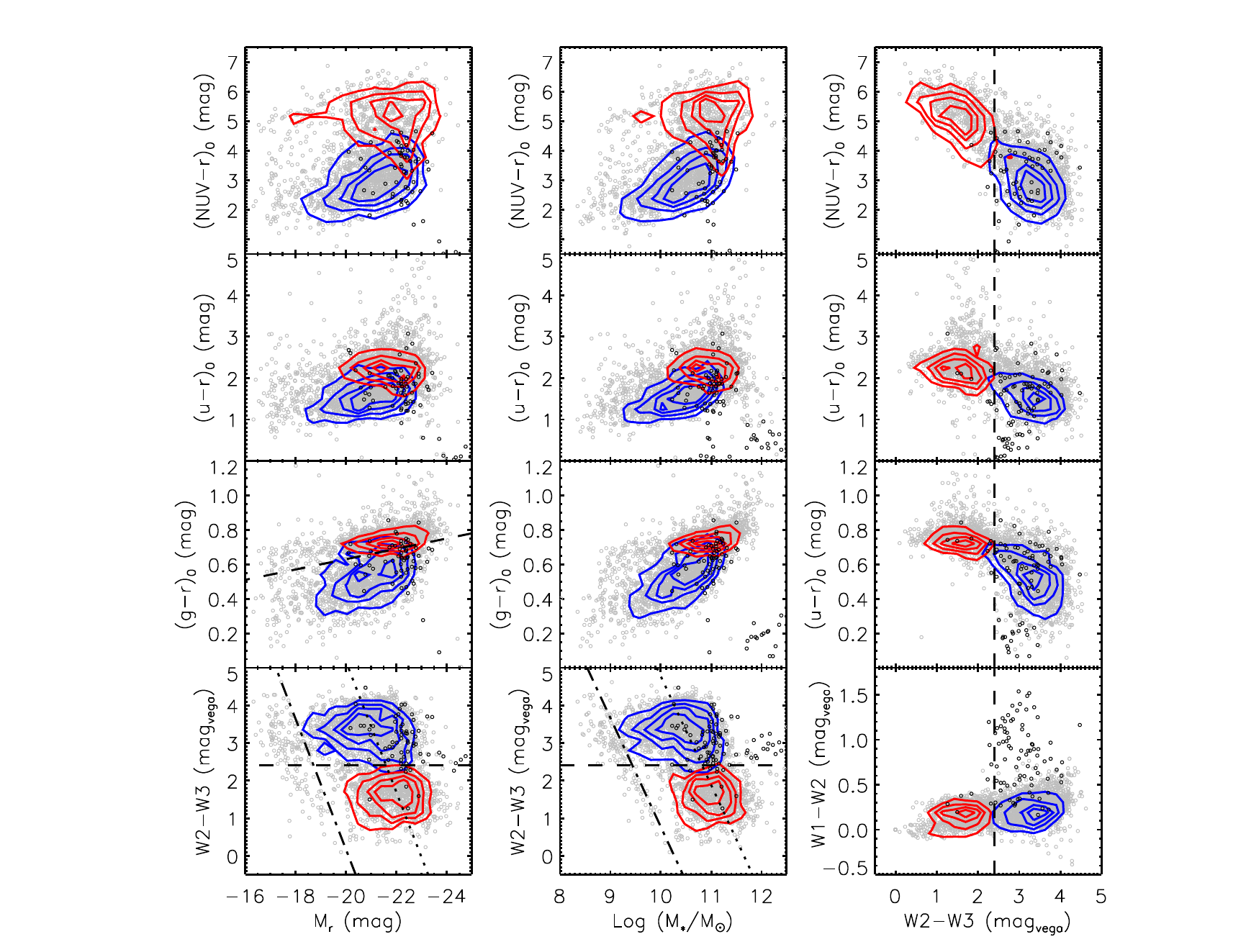}
\caption{The color-magnitude diagrams of $M_r$/$(NUV-r)_0$, $M_r$/$(u-r)_0$, $M_r$/$(g-r)_0$, $M_r$/$W2-W3$ (left), the mass-color diagrams of $M_*$/$(NUV-r)_0$, $M_*$/$(u-r)_0$, $M_*$/$(g-r)_0$, $M_*$/$W2-W3$ (middle), and the color-color diagrams of $W2-W3$/$(NUV-r)_0$, $W2-W3$/$(u-r)_0$, $W2-W3$/$(g-r)_0$, $W2-W3$/$W1-W2$ (right). The gray dots indicate each galaxy, while the red and blue contours represent the number densities of absorption and emission line galaxies, respectively. The dashed lines indicate the separation of MIR-detected absorption and emission line galaxies, except for the $M_r$/$(g-r)_0$ diagram in which it indicates the separation of the rest-frame bimodality of galaxies from \citet{Bernardi2010}. The double checked AGNs (spectra and $W2-W3$/$W1-W2$ colors) have been marked on each diagrams with black points. The dashed-dot lines separate the fainter paralleled sequence from the main population of galaxies (fitted by the dotted lines) in both $M_r$/$W2-W3$ and $M_*$/$W2-W3$ diagrams (see text for details). \label{all_cmd}}
\end{center}
\end{figure*}

Figure~\ref{all_cmd} shows the color-magnitude diagrams of $M_r$/$(NUV-r)_0$, $M_r$/$(u-r)_0$, $M_r$/$(g-r)_0$, $M_r$/$W2-W3$, the mass-color diagrams of $M_*$/$(NUV-r)_0$, $M_*$/$(u-r)_0$, $M_*$/$(g-r)_0$, $M_*$/$W2-W3$, and the color-color diagrams of $W2-W3$/$(NUV-r)_0$, $W2-W3$/$(u-r)_0$, $W2-W3$/$(g-r)_0$, $W2-W3$/$W1-W2$, respectively. For clarity, the contours of absorption (red) and emission line galaxies (blue) are shown on each diagram. The separation of the rest-frame bimodality of galaxies as 
\begin{equation}
(g-r)_0=0.63-0.03\times(M_{r}+20),
\end{equation}
has been added to the $M_r$/$(g-r)_0$ diagram \citep{Bernardi2010}. The double checked AGNs (spectra and $W2-W3$/$W1-W2$ colors) also have been marked on each diagram with black points. As the diagrams shown, the separation of absorption and emission line galaxies by using $W2-W3=2.4$ is distinct in all of the $W2-W3$ related panels compared to other color indexes. 

\begin{figure}
\begin{center}
\includegraphics[clip, trim=5cm 14cm 2cm 4cm, scale=0.65]{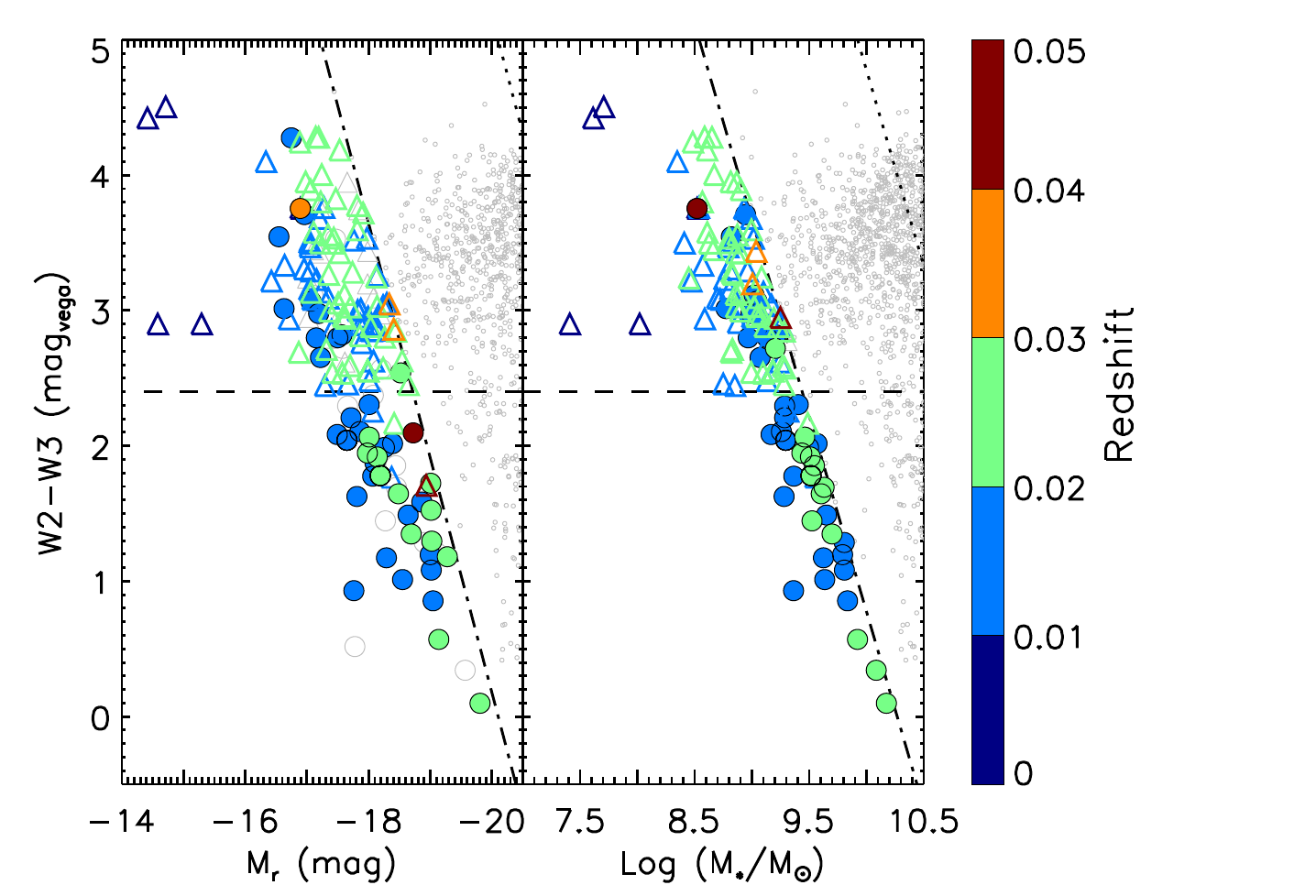}
\caption{Zoom-in regions of the $M_r$/$W2-W3$ and $M_*$/$W2-W3$ diagrams. The dashed-dot lines indicate the separation between the fainter paralleled sequence and the main population of galaxies. 127 common galaxies from both panels are color coded by redshift, while the distinct ones are shown as gray color. Four outliers on the left side of each panel are actually three galaxies, for which two outliers with $W2-W3\approx2.9$ are two HII regions in the same galaxy. \label{dwarf}}
\end{center}
\end{figure}

Meanwhile, a fainter sequence paralleled to the main population of galaxies has been witnessed both in $M_r$/$W2-W3$ and $M_*$/$W2-W3$ diagrams. This parallel sequence could be the population of luminous dwarf galaxies ($M_{r}\sim-18~mag$; \citealt{Wu2011}), which mimics the behavior of main population of galaxies. For better separation of the parallel sequence, the median values ($median_{M_{r}}$/$median_{M_{*}}$ and $median_{W2-W3}$) of absorption ($W2-W3\leq2.4$) and emission line galaxies ($W2-W3\geq2.4$) have been fitted with a linear function. The parallel sequences in the two diagrams have been simply separated as 
\begin{equation}
W2-W3=1.735\times M_{r}+39.882-5,
\end{equation}
and
\begin{equation}
W2-W3=-2.893\times M_{*}+33.703-4,
\end{equation}
where the last item in each formula (`5' and `4') indicates the offset from the linear fitting, respectively. There are 160 galaxies in the $M_r$/$W2-W3$ diagram and 132 galaxies in the $M_*$/$W2-W3$ diagram have been selected based on the criteria. The cross-matching between these two sets of galaxies indicates that there are 127 galaxies in common. Figure~\ref{dwarf} shows the zoom-in regions of the two diagrams with the common galaxies color coded by redshift, while the distinct ones shown as gray color. Four outliers on the left side of each panel turn out to be three galaxies, for which two outliers with $W2-W3\approx2.9$ are two HII regions in the same galaxy. Still, these three galaxies could be the candidates of very faint dwarf galaxies/low surface brightness galaxies inferred both from the diagrams and SDSS images. They have been labeled in our catalog. In addition, further SDSS image inspection of the 127 common galaxies shows that there is a major contamination ($\sim30\%$) from the edge-on/highly inclined galaxies and a small fraction ($\sim3\%$) of targets could be the HII regions misidentified as galaxies. Thus, follow up may be needed to confirm this parallel sequence.

\begin{figure*}
\begin{center}
\includegraphics[clip, trim=2cm 12cm 2cm 2cm, scale=0.57]{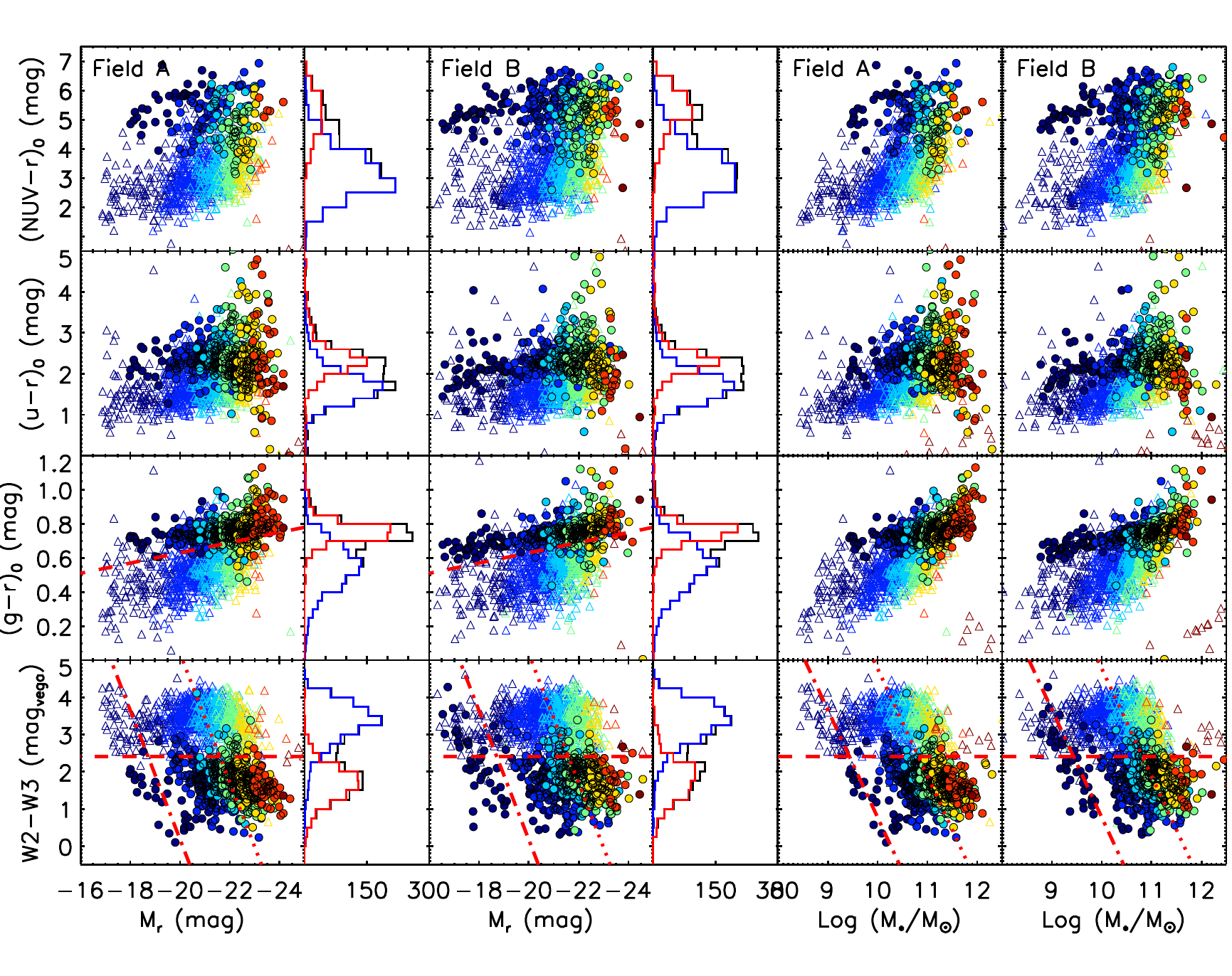}
\includegraphics[clip, trim=5.5cm 12cm 2.5cm 2cm, scale=0.57]{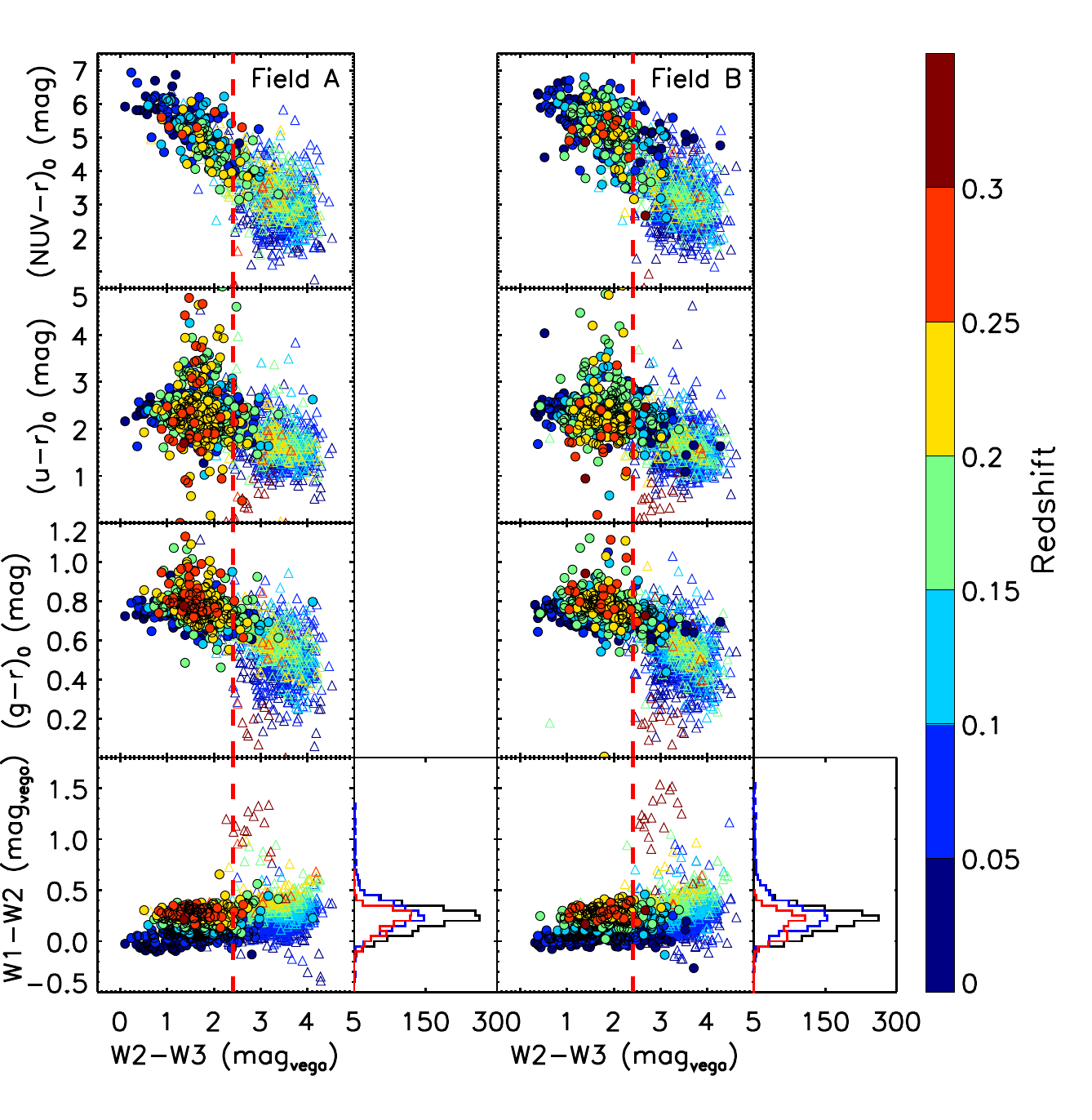}
\caption{Same as Fig~\ref{all_cmd} but for individual field. Each galaxy has been color coded by redshift. The histograms of each unique color index are shown on the right side of the panel. \label{cmd_individual}}
\end{center}
\end{figure*}

Alternatively, the distribution of individual field for each panel in Figure~\ref{all_cmd} also has been shown in Figure~\ref{cmd_individual}.

\section{Summary}

We present here a spectroscopic redshift catalog from the LAMOST Key Project, LaCoSSPAr, which is designed to observe all sources (Galactic and extra-galactic) by using repeating observations with a limiting magnitude of $r=18.1~mag$ in two $20~deg^2$ FOVs in the SGC. The project is mainly focusing on the completeness of LEGAS in the SGC, the deficiencies of source selection methods and the basic performance parameters of the LAMOST telescope. Two chosen FOVs represent the relatively lower (Field A) and higher (Field B) density of galaxies. Targets in the FOV are mainly consisted of stars, galaxies, quasars and u-band variables, as well as complementary HII regions.

We conduct the observation from September 2012 to January 2014. Field A contains 2519 galaxies and 13930 stars in which 2447 ($\sim97\%$) and 13278 ($\sim95\%$) of them have been observed ($97\%$) with 5 B- ($r=14.0\sim16.0~mag$) and 12 F-plates ($r=16.0\sim18.1~mag$), respectively. Field B contains 3104 galaxies and 11381 stars in which 2995 ($\sim97\%$) and 10224 ($\sim90\%$) of them have been observed with 6 B- and 5 F-plates, respectively. 

The raw data have been reduced with LAMOST 2D and 1D pipelines. An additional post-processing has been applied to LAMOST 1D spectrum to remove the majority of remaining sky background residuals. More than 10,000 spectra have been visually inspected to measure redshift by using combinations of different emission/absorption features with uncertainty of $\sigma_{z}/(1+z)<0.001$. All galaxies have been divided into two categories as emission and absorption line galaxies, depending on whether there are obvious emission lines appeared or not. 

In total, there are 1528 redshifts ($\sim62\%$ of observed galaxies; including 623 absorption and 905 emission line galaxies) have been measured in Field A with median value of $z=0.125$ and range from 0.001 to 1.598 (99\% at $z<0.3$). For Field B, there are 1570 redshifts ($\sim53\%$ of observed galaxies; including 569 absorption and 1001 emission line galaxies) have been measured with median value of $z=0.0952$ and range from 0.006 to 3.148 (99\% at $z<0.3$). The results show that it is possible to derive redshift from low SNR galaxies with our post-processing and visual inspection, since the $SNR_r$ is less than 5 for about 41\% and 24\% of galaxies in Field A and B, respectively. The comparison between our result and SDSS shows that, for 286 galaxies in Field A and 506 galaxies in Field B which both have LAMOST and SDSS spectroscopic redshift within a search radius of 1", all matched galaxies have almost identical redshift with $\sigma_{\Delta z}=0.0004$ and $\sigma_{\Delta z}=0.0002$ for Field A and B, respectively. Further visual inspection of SDSS image indicate that up to 1/4 of the input targets for a typical extra-galactic spectroscopic survey might be unreliable. The improvement of classification algorithm of the prior photometric survey could significantly improve the efficiency of the following spectroscopic survey which, may be important for the forthcoming large-scale spectroscopic surveys like 4MOST, MOONS, WEAVE, DESI, PFS, MSE and so on.

Our sample also has been cross-matched with GALEX and ALLWISE catalog to obtain the multi-wavelength data from FUV to MIR. Based on the analysis, it has been revealed that the majority of MIR-detected absorption (91.3\%) and emission line galaxies (93.3\%) can be well separated by an empirical criterion of $W2-W3=2.4$. Further confirmation has been made by using MPA/JHU value-added galaxy catalog for SDSS DR7, which also indicates that the inclusion of `non-LINER AGNs' and `low SNR LINER' in the sample of emission line galaxies will contaminate the separation. Meanwhile, a fainter sequence paralleled to the main population of galaxies has been witnessed both in $M_r$/$W2-W3$ and $M_*$/$W2-W3$ diagrams, which could be the population of luminous dwarf galaxies but also suffered from a major contamination ($\sim30\%$) of edge-on/highly inclined galaxies. Follow up may be needed to confirm this parallel sequence.

\section{Acknowledgments}

We would like to thank the staff of LAMOST at Xinglong Station for their excellent support during our observing runs.

This project is supported by the China Ministry of Science and Technology under the State Key Development Program for Basic Research (2014CB845705); the National Natural Science Foundation of China (Grant No.11733006, 11225316); the National Science Foundation for Young Scientists of China (Grant No.11403061, 11603058); National Key R\&D Program of China, 2017YFA0402704; the Strategic Priority Research Program ``The Emergence of Cosmological Structures'' of the Chinese Academy of Sciences (Grant No.XDB09000000); the Guoshoujing Telescope Spectroscopic Survey Key Projects.

Guoshoujing Telescope (the Large Sky Area Multi-Object Fiber Spectroscopic Telescope LAMOST) is a National Major Scientific Project built by the Chinese Academy of Sciences. Funding for the project has been provided by the National Development and Reform Commission. LAMOST is operated and managed by the National Astronomical Observatories, Chinese Academy of Sciences.

Funding for SDSS-III has been provided by the Alfred P. Sloan Foundation, the Participating Institutions, the National Science Foundation, and the U.S. Department of Energy Office of Science. The SDSS-III web site is http://www.sdss3.org/.

SDSS-III is managed by the Astrophysical Research Consortium for the Participating Institutions of the SDSS-III Collaboration including the University of Arizona, the Brazilian Participation Group, Brookhaven National Laboratory, Carnegie Mellon University, University of Florida, the French Participation Group, the German Participation Group, Harvard University, the Instituto de Astrofisica de Canarias, the Michigan State/Notre Dame/JINA Participation Group, Johns Hopkins University, Lawrence Berkeley National Laboratory, Max Planck Institute for Astrophysics, Max Planck Institute for Extraterrestrial Physics, New Mexico State University, New York University, Ohio State University, Pennsylvania State University, University of Portsmouth, Princeton University, the Spanish Participation Group, University of Tokyo, University of Utah, Vanderbilt University, University of Virginia, University of Washington, and Yale University.

This publication makes use of data products from the Wide-field Infrared Survey Explorer, which is a joint project of the University of California, Los Angeles, and the Jet Propulsion Laboratory/California Institute of Technology, funded by the National Aeronautics and Space Administration.

GALEX (Galaxy Evolution Explorer) is a NASA Small Explorer, launched in April 2003. We gratefully acknowledge NASA's support for construction, operation, and science analysis for the GALEX mission.


\begin{deluxetable}{cccc}
\tabletypesize{\scriptsize}
\tablecaption{Observational Information of LaCoSSPAr \label{tbl1}}
\tablewidth{0pt}
\tablehead{
	\colhead{Date} & \colhead{Plate Name} & \colhead{Exposure time} & \colhead{Dome Seeing} \\
	\colhead{(YYYY/MM/DD)} & & \colhead{(second)} & \colhead{(arcsec)}
}
\startdata
Field A &&&\\
\hline
20121013 & EG023131N032619F05 & 3$\times$1800 & 3.3 \\
20121113 & EG023131N032619F01 & 5$\times$1800 & 6.4 \\ 
20121114 & EG023131N032619F02 & 5$\times$1800 & 2.8 \\ 
20121209 & EG023131N032619F06 & 4$\times$1800 & 3.1 \\ 
20130105 & EG023131N032619F03 & 3$\times$1800 & 3.8 \\ 
20130106 & EG023131N032619F04 & 4$\times$1800 & 3.6 \\ 
20130108 & EG023131N032619F08 & 4$\times$1800 & 4.0 \\ 
20130109 & EG023131N032619F09 & 3$\times$1800 & 2.9 \\ 
20131002 & EG023131N032619F07 & 3$\times$1800 & 3.2 \\
20131002 & EG023131N032619B01 & 1500+1200+850 & 3.4 \\  
20131030 & EG023131N032619F0A & 3$\times$1800 & 2.8 \\
20131030 & EG023131N032619B02 & 3$\times$600 & 3.2 \\
20131201 & EG023131N032619F0B & 3$\times$1800 & 3.0 \\
20131201 & EG023131N032619B03 & 3$\times$600 & 3.6 \\
20131226 & EG023131N032619F0C & 3$\times$2400 & 3.9 \\
20140105 & EG023131N032619B04 & 3$\times$600 & 4.0 \\
20140121 & EG023131N032619B05 & 3$\times$1500+900 & 3.4 \\
\hline
Field B &&&\\
\hline
20121012 & EG012606S021203F03 & 2$\times$1800 & 3.0 \\ 
20121101 & EG012606S021203B03 & 2$\times$900  & 3.6 \\
20131025 & EG012606S021203F01 & 3$\times$1800 & 2.6 \\
20131029 & EG012606S021203B01 & 3$\times$600  & 3.8 \\
20131103 & EG012606S021203F02 & 3$\times$1800 & 3.8 \\
20131107 & EG012606S021203F04 & 3$\times$1800 & 3.7 \\
20131203 & EG012606S021203B02 & 3$\times$600  & 3.4 \\
20131205 & EG012606S021203F05 & 3$\times$1800 & 3.1 \\
20131227 & EG012606S021203B04 & 3$\times$600  & 4.2 \\
20140103 & EG012606S021203B05 & 3$\times$600  & 3.0 \\
20140109 & EG012606S021203B06 & 3$\times$600  & 4.2 \\
\enddata
\end{deluxetable}
\clearpage

\begin{deluxetable}{ccccccccccc}
\tabletypesize{\scriptsize}
\tablecaption{The Spectroscopic Redshift Catalog of LaCoSSPAr Field A\label{tbl2}}
\tablewidth{0pt}
\tablehead{
\colhead{ID} & \colhead{RA} & \colhead{Dec} & \colhead{Type\tablenotemark{a}} & \colhead{Redshift} & \colhead{FUVmag} & \colhead{$e\_FUVmag$} & \colhead{...} & \colhead{$rest\_{imag}$} & \colhead{$rest\_{zmag}$} & \colhead{Comments\tablenotemark{b}}
}
\startdata
   1 &  36.856689  &   2.105210 &     e &  0.0098 &  -999.000 &  -999.000  &  ...  &  16.214  &   16.328 &     \\       
   2 &  37.634655  &   3.022465 &     e &  0.0146 &    18.665 &     0.109  &  ...  &  15.785  &   15.868 &     \\       
   3 &  39.556831  &   1.858582 &     e &  0.0190 &    19.908 &     0.227  &  ...  &  16.513  &   16.457 &     \\       
   4 &  39.658295  &   2.053094 &     e &  0.0195 &  -999.000 &  -999.000  &  ...  &  16.310  &   16.077 &     \\       
   5 &  39.385441  &   2.421656 &     e &  0.0197 &    21.246 &     0.368  &  ...  &  16.535  &   16.328 &     \\       
...  &   ...       &  ...       &  ...  &  ...    &  ...      &  ...       &  ...  &  ...     &   ...    &  ...\\
\enddata
\tablenotetext{a}{Galaxy type: `a' for absorption line galaxy, `e' for emission line galaxy.}
\tablenotetext{b}{Candidates of peculiar targets: BL: broadline; HII: HII region; LN: LINER; OL: overlapped galaxies/star-galaxy; QSO: quasar; Sy1: Seyfert1; Sy2: Seyfert2. The "?" indicates the tentative identification.}
\tablecomments{This table is available in its entirety in a machine-readable form in the online journal. A portion is shown here for guidance regarding its form and content.}
\end{deluxetable}

\begin{deluxetable}{ccccccccccc}
\tabletypesize{\scriptsize}
\tablecaption{The Spectroscopic Redshift Catalog of LaCoSSPAr Field B\label{tbl3}}
\tablewidth{0pt}
\tablehead{
\colhead{ID} & \colhead{RA} & \colhead{Dec} & \colhead{Type\tablenotemark{a}} & \colhead{Redshift} & \colhead{FUVmag} & \colhead{$e\_FUVmag$} & \colhead{...} & \colhead{$rest\_{imag}$} & \colhead{$rest\_{zmag}$} & \colhead{Comments\tablenotemark{b}}
}
\startdata
   1    & 19.140558   & -1.703767   &   e   & 0.0060  & -999.000  & -999.000   &    ...      &  16.800     & 16.905  &   HII/VFD?\\
   2    & 19.142012   & -1.703741   &   e   & 0.0060  & -999.000  & -999.000   &    ...      &  18.211     & 25.697  &   HII/VFD?\\
   3    & 22.656858   & -3.733585   &   e   & 0.0066  &   20.354  &    0.193   &    ...      &  17.732     & 17.631  &       VFD?\\
   4    & 22.224884   & -1.699854   &   e   & 0.0066  &   19.807  &    0.177   &    ...      &  17.306     & 18.361  &       VFD?\\
   5    & 22.801865   & -0.610780   &   e   & 0.0155  &   20.793  &    0.166   &    ...      &  16.665     & 16.349  &           \\ 
   ...  &   ...       &  ...        &  ...  &  ...    &  ...      &  ...       &    ...      &  ...        &  ...    &  ...      \\
\enddata
\tablenotetext{a}{Galaxy type: `a' for absorption line galaxy, `e' for emission line galaxy.}
\tablenotetext{b}{Candidates of peculiar targets: BL: broadline; HII: HII region; LN: LINER; OL: overlapped galaxies/star-galaxy; QSO: quasar; Sy1: Seyfert1; Sy2: Seyfert2; VFD: very faint dwarf galaxies. The "?" indicates the tentative identification.}
\tablecomments{This table is available in its entirety in a machine-readable form in the online journal. A portion is shown here for guidance regarding its form and content.}
\end{deluxetable}

\end{CJK*}

\end{document}